\documentclass[superscriptaddress,twocolumn,prb,aps,showpacs,floatfix]{revtex4}

\usepackage{epsfig}
\usepackage{epsf}
\usepackage{graphicx}
\usepackage{amsmath}
\usepackage{booktabs}  
\usepackage{dcolumn}
\usepackage{rotating}
\usepackage{bm}        
\usepackage{bbm}       
\newcommand{\be}{\begin{equation}}
\newcommand{\ee}{\end{equation}}
\newcommand{\bea}{\begin{eqnarray}}
\newcommand{\eea}{\end{eqnarray}}
\newcommand{\ba}{\begin{array}}
\newcommand{\ea}{\end{array}}
\newcommand{\mint}[1]{\int\! \D^{3} #1 \, }
\newcommand{\mdint}[2]{\mint{#1}\!\!\!\mint{#2}}

\newcommand{\I}{{\rm i}}
\newcommand{\D}{{\rm d}}
\newcommand{\E}{{\rm e}}

\newcommand{\bG}{\bm{G}}

\newcommand{\bk}{{\bm{k}}}

\newcommand{\bq}{\bm{q}}
\newcommand{\br}{\bm{r}}
\newcommand{\bR}{\bm{R}}

\newcommand{\bv}{\bm{v}}

\newcommand{\bek}[1]{\left(\frac{\beta}{2}{#1}\right)}

\newcommand{\KS}{{\rm s}}

\newcommand{\nk}{{n\bk}}
\newcommand{\nkp}{{n'\bk'}}

\newcommand{\xik}{\xi_\nk}
\newcommand{\xikp}{\xi_\nkp}
\newcommand{\Ek}{E_\nk}

\begin{document}


\newcommand{\PI}{I}
\newcommand{\LDAzero}{TF-FE}
\newcommand{\LDAone}{TF-SK}
\newcommand{\DFT}{TF-ME}
\newcommand{\FH}{{\cal K}^\text{el}}
\newcommand{\Zdef}{{\cal Z}}
\newcommand{\Kdef}{{\cal K}^\text{ph}}

\unitlength1cm

\title{Ab-initio theory of superconductivity -- II: Application to elemental metals}

\author{M.\,A.\,L.~Marques}
\affiliation{Institut f{\"u}r Theoretische Physik, Freie Universit{\"a}t Berlin, Arnimallee 14, D-14195 Berlin, Germany} 
\affiliation{Institut~f\"ur Theoretische Physik, Universit\"at 
W\"urzburg, Am Hubland, D-97074 W\"urzburg, Germany}

\author{M.~L{\"u}ders}
\affiliation{Daresbury Laboratory, Warrington WA4 4AD, United Kingdom }
\affiliation{Institut~f\"ur Theoretische Physik, Universit\"at 
W\"urzburg, Am Hubland, D-97074 W\"urzburg, Germany}

\author{N.\,N.~Lathiotakis}
\affiliation{Institut f{\"u}r Theoretische Physik, Freie Universit{\"a}t Berlin, Arnimallee 14, D-14195 Berlin, Germany} 
\affiliation{Institut~f\"ur Theoretische Physik, Universit\"at 
W\"urzburg, Am Hubland, D-97074 W\"urzburg, Germany}

\author{G.~Profeta}
\affiliation{CASTI - Istituto Nazionale Fisica della Materia
(INFM) and Dipartimento di Fisica, Universit\`a degli studi dell'Aquila,
I-67010 Coppito (L'Aquila) Italy}

\author{A.~Floris}  
\affiliation{Institut f{\"u}r Theoretische Physik, Freie Universit{\"a}t Berlin, Arnimallee 14, D-14195 Berlin, Germany} 
\affiliation{INFM SLACS, Sardinian Laboratory for Computational Materials Science and
Dipartimento di Scienze Fisiche, Universit\`a degli Studi di Cagliari,
S.P. Monserrato-Sestu km 0.700, I--09124 Monserrato (Cagliari), Italy}

\author{L.~Fast}
\affiliation{SP Swedish National Testing and Research
Institute, P.O.B. 857, S-501 15 Bor{\aa}s, Sweden}
\affiliation{Institut~f\"ur Theoretische Physik, Universit\"at 
W\"urzburg, Am Hubland, D-97074 W\"urzburg, Germany}

\author{A.~Continenza}
\affiliation{CASTI - Istituto Nazionale Fisica della Materia
(INFM) and Dipartimento di Fisica, Universit\`a degli studi dell'Aquila,
I-67010 Coppito (L'Aquila) Italy}

\author{E.\,K.\,U.~Gross}
\affiliation{Institut f{\"u}r Theoretische Physik, Freie Universit{\"a}t Berlin, Arnimallee 14, D-14195 Berlin, Germany} 
\affiliation{Institut~f\"ur Theoretische Physik, Universit\"at 
W\"urzburg, Am Hubland, D-97074 W\"urzburg, Germany}

\author{S.~Massidda}
\altaffiliation{Also at LAMIA-INFM, Genova, Italy}
\affiliation{INFM SLACS, Sardinian Laboratory for Computational Materials Science and
Dipartimento di Scienze Fisiche, Universit\`a degli Studi di Cagliari,
S.P. Monserrato-Sestu km 0.700, I--09124 Monserrato (Cagliari), Italy}

\begin{abstract}  
The density functional theory for superconductors developed in the preceding article
is applied to the calculation of superconducting properties of several elemental metals.
In particular, we present results for the transition temperature, for the gap at zero temperature, 
and for thermodynamic properties like the specific heat. We obtain an unprecedented agreement 
with experimental results. Superconductors both with strong and weak electron-phonon coupling 
are equally well described. This demonstrates that, as far as conventional superconductivity is 
concerned, the first-principles prediction of superconducting properties is feasible.
\end{abstract}

\maketitle 

\section{Introduction}
\label{secintro}

The recent discovery of superconductivity at around 40\,K in MgB$_2$~\cite{mgb2_dis} has renewed the attention of 
the scientific community on this field. MgB$_2$ is just one in a long list of materials that
are found to be superconducting. This list includes several elemental metals, heavy fermion
compounds, high-$T_\text{c}$ ceramics~\cite{hightc}, fullerenes doped with alkali atoms~\cite{fulle}, etc. 
The mechanism responsible for superconductivity can have different origins. For example, in the 
elemental metals, in the fullerenes~\cite{fulle}, and also in MgB$_2$~\cite{mgb2-theor,MgB2-PRL} 
the electron-phonon interaction 
is the responsible for the binding of the Cooper pairs. This situation is usually
referred to as ``conventional superconductivity''. On the other hand, it is generally
believed that the very high transition temperatures exhibited by the high-$T_\text{c}$'s
are (at least partly) due to Coulombic effects. Following the remarkable experimental discoveries of the
last years, there were numerous theoretical developments, that have greatly improved our
description and understanding of superconductivity. However, the prediction of 
material-specific properties of superconductors still remains one of the great challenges 
of modern condensed-matter theory.

In this work, we present an ab-initio theory to describe the superconducting state. It is
based on a density functional formulation, and is capable of describing both weakly and strongly
coupled superconductors. This is achieved by treating the electron-phonon and Coulomb
interactions on the same footing. The main equation of this theory resembles the gap equation
of the theory of Bardeen, Cooper and Schrieffer\cite{bcs}. It is, however, free of any adjustable parameter
and contains effects originating from the retarded nature of the electron-phonon interaction.
The theoretical foundations of our approach are presented in the preceding
paper~\cite{first}, which will henceforth be referred to as I.
As in ordinary density functional theory (DFT)~\cite{HK,KS,Grossbook}, the complexities of the many-body problem are 
included in an exchange-correlation functional. In I we use Kohn-Sham perturbation theory~\cite{GoerlingLevy} to 
derive several approximations for this quantity. In the present paper we describe the implementation 
of our theory, and its application to the calculation of superconducting properties of elemental metals.
The systems under consideration range from weak-coupling (Mo, Al, Ta) to strong-coupling (Nb, Pb) 
superconductors. By studying these well-known systems, we illustrate the usefulness and 
accuracy of DFT for superconductors. Furthermore, our results serve as a justification for the 
choices and approximations made in I. Further applications of our approach to more complex systems 
like MgB$_2$~\cite{MgB2-PRL} or solids under pressure will be presented in separate publications.

This paper is organized as follows: In Sect.~\ref{sec:method} we give a brief 
summary of the theoretical foundations of our work. The exchange-correlation functionals
are described in Section~\ref{sec:functionals}. We present three different levels of 
approximation: functionals that retain the full dependence on the wave-vector, energy
averaged functionals, and hybrid schemes. We proceed with an account of the computational 
details of our numerical implementation. In Sect.~\ref{sec:res} we present and 
discuss numerical results obtained for the elemental metals. These results include
calculations of the transition temperature $T_\text{c}$, the gap at zero temperature 
$\Delta_0$, and thermodynamic properties like the specific heat.
The last section is devoted to the conclusions.

\section{Method}
\label{sec:method}

In this section we give a brief account of the theoretical foundations of DFT
for the superconducting state. For an in depth description of this theory, we refer the reader to \PI.

A correct description of the superconducting state has to include the effects of the electron-electron 
and electron-phonon interactions. DFT for superconductors is a theory designed to treat on the same footing both
electronic correlations and the electron-phonon coupling. To achieve this unified description, 
we start with the full electron-nuclear Hamiltonian. A multi-component DFT~\cite{kreibich} is then established using 
a set of three densities: i)~the normal electronic density $n(\br)$;
ii)~the anomalous density $\chi(\br,\br')$, which is the order parameter of the superconducting state;
and iii)~the diagonal of the nuclear density matrix $\Gamma(\underline{\bR})$, where $\underline{\bR}$
is a shorthand for the $N$ nuclear coordinates $\{\bR_1,\bR_2,\cdots,\bR_N\}$.
An extension of the Hohenberg-Kohn theorem~\cite{HK,mermin} 
guarantees a one-to-one correspondence between the set of the densities 
$\{n(\br),\chi(\br,\br'),\Gamma(\underline{\bR})\}$ in thermal equilibrium and the set of their conjugate 
potentials. As a consequence, all observables are functionals of this set of densities.

We then construct a Kohn-Sham system~\cite{KS} composed of non-interacting (but superconducting) electrons
and nuclei. The latter interact with each other through an $N$-body potential, but do not interact
with the electrons. The Kohn-Sham system is chosen such that the Kohn-Sham densities are equal to
the densities of the interacting system. 
The Euler-Lagrange equations for the Kohn-Sham system lead to a set of three coupled equations, 
one of which describing the nuclear degrees of freedom, and the other two describing the electrons. 
These three equations have to be solved self-consistently. 
The nuclear equation describes a set of $N$ nuclei under the influence of an effective $N$-body potential 
$v^\text{n}_\KS[n,\chi,\Gamma] (\underline{\bR})$, and has the same structure as the usual
Born-Oppenheimer equation for the nuclei. In this article we are interested in solids at relatively low 
temperature, where the nuclei perform small oscillations around their equilibrium positions. 
In this case, we can expand $v_{\KS}^\text{n}[n,\chi,\Gamma](\underline{\bR})$ in a Taylor series around the 
equilibrium positions, and transform the nuclear degrees of freedom into collective (phonon) coordinates.
The Kohn-Sham electrons obey a system of two coupled equations
\begin{subequations}
\label{KS-BdG}
\bea
  \left[ -\frac{\nabla^2}{2} + v^\text{e}_\KS(\br) - \mu \right] u_\nk(\br)
  &+& \mint{r'} \Delta_\KS(\br,\br') v_\nk(\br') \nonumber \\ 
  &=& E_\nk \, u_\nk(\br) \\
  - \left[ -\frac{\nabla^2}{2} + v^\text{e}_\KS(\br) - \mu \right] v_\nk(\br)
  &+& \mint{r'} \Delta^*_\KS(\br,\br') u_\nk(\br') \nonumber \\
  &=& E_\nk \, v_\nk(\br)
  \,.
\eea
\end{subequations}
Note that for a periodic solid, we can classify the corresponding solutions according
to the symmetry of the translational group of the crystal.  Thus we label the eigenstates
with a principal quantum number $n$ and a wave-vector quantum number $\bk$.
The Eqs.~(\ref{KS-BdG}) have the same structure as the Bogoliubov-de~Gennes equations,
and describe a system of non-interacting, superconducting, electrons moving
under the influence of the effective potential $v^\text{e}_\KS(\br)$ and the
effective pairing field $\Delta_\KS(\br,\br')$. As it is usual in DFT, the effective
potentials $\{v^\text{e}_\KS, \Delta_\KS, v_{\KS}^\text{n}\}$ are functionals of
the set of densities $\{n, \chi, \Gamma\}$ and include Hartree and exchange-correlation
contributions. These latter terms, the exchange-correlation potentials, are defined
as functional derivatives with respect to the densities of the exchange-correlation 
contribution to the free energy $F_\text{xc}$.

The full self-consistent solution of the three Kohn-Sham equations is a highly demanding task.
For the time being, we resort to some additional approximations 
in order to simplify the problem. First, we neglect the dependence of the nuclear potential 
$v_{\KS}^\text{n}$ on $\chi$, which amounts
to neglecting the effect of the superconducting pair potential on the phonon dispersion. 
This effect has been measured experimentally~\cite{ph1}, and it turns out to be 
quite small. Furthermore, we assume that the nuclear potential $v_{\KS}^\text{n}$ is well approximated
by the Born-Oppenheimer potential. Within these approximations, the phonon frequencies and 
the electron-phonon coupling constants can be calculated using standard density functional 
linear-response methods~\cite{LRT}.

A direct solution of the Kohn-Sham Bogoliubov-de Gennes equations~(\ref{KS-BdG})~\cite{balazs} is faced 
with the problem that one needs extremely high accuracy to resolve the superconducting energy scale,
which is about three orders of magnitude smaller than typical electronic energies. 
At the same time, one has to cover the whole energy range of the electronic band structure.
The problem can be simplified through the decoupling approximation~\cite{stegross}.
First we assume that $v^\text{e}_\KS(\br)$ does not depend significantly on the anomalous density 
$\chi$. In fact, we expect that corrections to $v^\text{e}_\KS(\br)$ are of the order of $|\chi|^2$ and 
therefore much smaller than the typical electronic energy scale. The normal-state Kohn-Sham 
eigenfunctions and eigenenergies can then be used to construct approximations for the eigenstates 
of the superconducting phase
\be
  \label{eq:decapprox}
  u_\nk(\br) \approx u_\nk \varphi_\nk(\br)
  \quad ; \quad 
  v_\nk(\br) \approx v_\nk \varphi_\nk(\br)
  \,.
\ee
The $\varphi_\nk$'s are the solutions of the normal-state Kohn-Sham equations 
for band $n$ and wave vector $\bk$ and can thus be calculated using standard electronic 
structure methods. If $N_\text{b}$ is the number of bands, the decoupling approximation
transforms, for every $\bk$ point in the Brillouin zone, the $2N_\text{b}\times2N_\text{b}$
eigenvalue problem given by Eq.~(\ref{KS-BdG}) into $N_\text{b}(2\times2)$ secular equations.
The Kohn-Sham eigenenergies then become
\be
  \label{eq:ek}
  \Ek = \sqrt{\xi_\nk^2+|\Delta_\nk|^2}
  \,,
\ee
where $\xi_\nk$ are the normal-state Kohn-Sham eigenvalues measured relative to the
chemical potential $\xi_\nk = \epsilon_\nk - \mu$. The matrix elements $\Delta_{\nk}$, which
are a central quantity in our formalism, are defined as
\be
  \label{eq:thegap}
  \Delta_\nk=\mdint{r}{r'} \varphi_\nk^*(\br)\Delta_\KS(\br,\br')\varphi_\nk(\br')
  \,.
\ee

Within the decoupling approximation, $\Delta_\nk$ is determined
by the integral equation
\be
  \label{eq:gap}
  \Delta_\nk = - {\cal Z}_\nk \Delta_\nk -\frac{1}{2}\sum_\nkp
  {\cal K}_{\nk,\nkp} \frac{\tanh\left(\frac{\beta}{2}E_\nkp\right)}{E_\nkp}
  \Delta_\nkp
  \,,
\ee
where $\beta$ is the inverse temperature. The quantities ${\cal Z}_\nk$ and ${\cal K}_{\nk,\nkp}$
are functionals of the pair potential $\Delta_\nk$ and of the chemical potential $\mu$.
We note that, although the gap equation~(\ref{eq:gap}) is static (i.e. it does not depend
explicitly on the frequency), it includes retardation effects through the
functionals ${\cal Z}_\nk$ and ${\cal K}_{\nk,\nkp}$. We will come back to this point
later in the discussion of our results.

It is possible to view the decoupling approximation from a different perspective:
It can be shown that the matrix elements of the pair potential $\Delta_\KS(\br,\br')$
are diagonal with respect to all symmetry-related quantum numbers, in particular the
Bloch wave-vector $\bk$, but also labels of the point-group symmetry (which usually are
not explicitly given).
The decoupling approximation amounts to neglecting 
the matrix elements which are off-diagonal with respect to the band index $n$.
For a given $\bk$-point in the Brillouin-zone, the corresponding states are in general
energetically far apart, which justifies the neglect of these matrix elements.
The only situation where this approximation may break down is when two bands of the same symmetry 
cross in the vicinity of the Fermi surface. We note in passing that the decoupling approximation is 
(trivially) exact for the uniform superconducting electron gas. 

The decoupling approximation can further be justified by the fact
that a perturbative treatment of the neglected off-diagonal terms
does not give any contribution in first order.\cite{stefan}  To see this, we
consider the neglected part of the pair-potential
\be
\tilde{\Delta}(\br,\br') = \sum_{\bk, n \neq n'} \varphi_{\nk}(\br)
\tilde{\Delta}_{\bk,n n'} \varphi^{*}_{\nkp}(\br')
\ee
and treat it by perturbation theory. The first-order correction to the
eigenenergies $E_{\nk}$ is given by
\be
\delta E^{(1)}_{\nk} = \mdint{r}{r'} u^{*}_{\nk}(\br) \tilde{\Delta}(\br,\br') v_{\nk}(\br')
+ h.c.
\ee
Inserting the eigenfunctions $u_{\nk}(\br)$ and $v_{\nk}(\br)$ within the
decoupling approximation, as given by eq.~(\ref{eq:decapprox}), and using
the orthogonality if the Bloch wave-functions,
$\mint{r} \varphi^*_{\nk}(\br) \varphi_{\nkp}(\br) = \delta_{n n'}$,
one finds
\begin{eqnarray}
\delta E^{(1)}_{\nk} &=&  u^{*}_{\nk} v_{\nk} \sum_{n_1 \neq n_2}
\tilde{\Delta}_{\bk,n_1 n_2} (\br')
\, \delta_{n n_1} \delta_{n_2 n}  + h.c. \,, \nonumber \\
&=& 0 \,.
\end{eqnarray}

Eq.~(\ref{eq:ek}) allows us to interpret $2\Delta_\nk$ as the 
superconducting gap. As a matter of principle, there is no guarantee that this 
gap reproduces the true gap of the superconductor: Even with the 
exact exchange-correlation functional, the Kohn-Sham system need not reproduce 
the true spectrum of the fully interacting system.  In semiconductors and 
insulators, the fundamental gap is given by the Kohn-Sham gap plus the 
discontinuity of the xc potential with respect to the particle number.
Standard functionals such as LDA and the GGAs are continuous with respect
to the particle number and, therefore, cannot reproduce the discontinuity.
In the superconducting case, the appearance of a discontinuity is not
expected because we are not working with fixed particle number in the first
place.  This fact alone does, of course, not prove that, for superconductors,
the Kohn-Sham gap resulting from the exact xc functional would be identical with
the true gap.  This question remains the subject of future investigations.

\section{Functionals}
\label{sec:functionals}
\subsection{$\bk$-dependent functionals}

In this work we use the partially linearized versions for the functionals ${\cal Z}_\nk$ and 
${\cal K}_{\nk,\nkp}$ proposed in \PI. The contributions stemming from the electron-phonon 
interaction are obtained with the help of Kohn-Sham perturbation theory. There are two
terms: i)~the first is non-diagonal
\begin{multline}
  \label{eq:kph_ij}
  {\cal K}^{\rm ph}_{\nk,\nkp} = \frac{2}{\tanh\bek{\xi_\nk}\tanh\bek{\xi_\nkp}}
  \sum_{\lambda,\bq} \left|g_{\lambda,\bq}^{\nk,\nkp}\right|^2 
   \\ \times \left[
    I(\xi_\nk,\xi_\nkp,\Omega_{\lambda,\bq}) - I(\xi_\nk,-\xi_\nkp,\Omega_{\lambda,\bq})
  \right]
  \,,
\end{multline}
where $g_{\lambda,\bq}^{\nk,\nkp}$ are the electron-phonon coupling constants and the function $I$ is
defined as
\begin{multline}
  \label{ikernel}
  I(\xi,\xi',\Omega) =
  f_\beta(\xi) \, f_\beta(\xi') \, n_\beta(\Omega)  \\
  \times
  \left[ \frac{\E^{\beta \xi} - \E^{\beta (\xi' + \Omega)}}{\xi - \xi' - \Omega}
  -  \frac{\E^{\beta \xi'} - \E^{\beta (\xi + \Omega)}}{\xi - \xi' + \Omega} 
  \right]
  \,.
\end{multline}
In the previous expression $f_\beta$ and $n_\beta$ are the Fermi-Dirac and Bose-Einstein
distributions; ii) The second contribution is diagonal in $\nk$ and reads
\begin{multline}
\label{eq:Z_ph}
  {\cal Z}^\text{ph}_\nk = \frac{1}{\tanh\left(\frac{\beta}{2} \xi_\nk\right)}
  \sum_\nkp \sum_{\lambda,\bq} \left| g^{\nk,\nkp}_{\lambda,\bq}\right|^2 \\
  \left[J(\xi_\nk,\xi_\nkp,\Omega_{\lambda,\bq})+J(\xi_\nk,-\xi_\nkp,\Omega_{\lambda,\bq})\right]
  \,,
\end{multline}
where the function $J$ is defined by
\be
  \label{jkernel}
  J(\xi,\xi',\Omega) = \tilde J(\xi,\xi',\Omega) - \tilde J(\xi,\xi',-\Omega)
  \,.
\ee
Finally we have
\begin{multline}
  \tilde J(\xi,\xi',\Omega) = - \frac{f_\beta(\xi) + n_\beta(\Omega)}{\xi-\xi'-\Omega} \\
  \times \left[
  \frac{f_\beta(\xi')-f_\beta(\xi-\Omega)}{\xi-\xi'-\Omega}
  -\beta f_\beta(\xi-\Omega)f_\beta(-\xi'+\Omega)
  \right]\,.
\end{multline}
On the other hand, the Coulomb interaction leads to the term
\be
  \label{fxcA}
  {\cal K}^\text{TF-ME}_{\nk,\nkp}=v^\text{TF}_{\nk,\nkp}
  \,,
\ee
with the definition
\begin{multline}
  \label{eq:vTF_ij}
  v^\text{TF}_{\nk,\nkp} = \mdint{r}{r'} v^{\rm TF}(\br-\br') \\
  \varphi^*_\nk(\br)\varphi_\nk(\br') \varphi_\nkp(\br)\varphi^*_\nkp(\br')
  \,.
\end{multline}
The electronic contribution is written in terms of the matrix elements (ME) of the 
screened Coulomb potential, as the use of the bare potential would be unrealistic. 
In this work we use a very simple model for the screening, namely the Thomas-Fermi (TF)
model
\be
  v^{\rm TF}(\br-\br') = \frac{e^{-k_{\text{TF}} |\br-\br'| }}{|\br - \br'|}
  \,,
\ee
with the Thomas-Fermi screening length, $k_{\rm TF}$, given by
\be
  \label{eq:TD_def}
  k^2_{\rm TF} = 4 \pi N(0) \,.
\ee
Finally, $N(0)$ denotes the total density of states at the Fermi level. Within this approach, 
the ME are calculated using the Bloch functions of the real material. In this basis, the ME 
read (using standard notation)
\begin{multline}
  \label{melocal}
  v^\text{TF}_{\nk,\nkp} = \frac{1}{\cal V}
  \sum_{\bG}\frac{4\pi}{|\bk - \bk' +\bG|^2+ k_\text{TF}^2} \\
  \times \left| \langle \nkp|\E^{-\I(\bk - \bk' +\bG)\cdot \br}|\nk \rangle \right|^2
  \,,
\end{multline}
where ${\cal V}$ is the volume of the unit cell. 

\subsection{Energy averaged functionals}
\label{numsec}

As discussed in the previous section, to obtain the gap function $\Delta_\nk$,
we solve the gap equation~(\ref{eq:gap}) with the functionals given by Eqs.~(\ref{eq:kph_ij}) and 
(\ref{eq:Z_ph}) for the electron-phonon interaction, and by Eq.~(\ref{eq:vTF_ij}) for the Coulomb repulsion.
The inputs required for such calculation are the electron-phonon coupling constants
$g^{\nk,\nkp}_{\lambda,\bq}$, and the normal-state Kohn-Sham eigenenergies
$\xi_\nk$ and eigenfunctions $\varphi_\nk$. The coupling constants can
be obtained from linear response~\cite{LRT} DFT calculations, while 
the eigenstates are the basic output of any standard DFT code.

It is possible, however, to further simplify the solution of
Eq.~(\ref{eq:gap}).  Very often, in the context of Eliashberg
theory~\cite{eliash}, one neglects the gap anisotropy over the Fermi
surface.  We can apply the same approximation in our context (see
Ref.~\onlinecite{martin} for further details). We assume that the pair
potential is constant on iso-energy surfaces, i.e. it depends on the energies only:
\begin{equation}
\Delta_{\nk} = \Delta(\xi_{\nk}) \,.
\end{equation}
We then insert the unity $1=\int\!\!\D\xi\; \delta(\xi-\xikp)$ under
the $\nkp$ summation in the right-hand side of Eq.(\ref{eq:gap}),
multiply the whole equation with the factor $\delta(\xi - \xi_{\nk})$
and perform the summation over all $\nk$. Finally we divide the
resulting equation by the density of states at the energy $\xi$,
$N(\xi)=\sum_\nk\delta(\xi-\xi_\nk)$.  This yields the gap equation in
energy space, which now is only an one-dimensional integral equation:
\begin{multline}
  \label{eq:gap_E}
  \Delta(\xi) = -{\cal Z}(\xi) \Delta(\xi)\\ - \frac{1}{2}\int_{-\mu}^{\infty} d\xi' N(\xi')
  {\cal K}(\xi,\xi') \frac{\tanh\left(\frac{\beta}{2}E'\right)}{E'} \Delta(\xi')
  \,.
\end{multline}
The energy-averaged functions ${\cal K}(\xi,\xi')$ and ${\cal Z}(\xi)$ are defined as:
\begin{equation}
\label{eq:K-ave}
{\cal K}(\xi,\xi') = \frac{1}{N(\xi) N(\xi')} \!
\sum_{\nk,\nkp} \! \! \delta(\xi-\xi_\nk)
\delta(\xi'-\xi_\nkp) K_{\nk,\nkp}  \,,
\end{equation}
\begin{equation}
\label{eq:Z-ave}
{\cal Z}(\xi) =  \frac{1}{N(\xi)} \sum_\nk \! \delta(\xi-\xi_\nk) Z_\nk  \,.
\end{equation}
We should mention that in performing the average over iso-energetic
surfaces we assumed an $s$-wave pairing field. It is straightforward
to devise similar averaging procedures for pairing fields of different
symmetry.

The phononic contributions to the averaged functionals read
\begin{multline}
  \label{fxcCave}
  {\cal K}^\text{ph}(\xi,\xi') = 
  \frac{2}{\tanh\left(\frac{\beta}{2}\xi\right)\tanh\left(\frac{\beta}{2}\xi'\right)}\frac{1}{N(0)}
  \int\!\!\D\Omega\; \alpha^2F(\Omega) \\  \times
  \left[I(\xi,\xi',\Omega) - I(\xi,-\xi',\Omega)\right]
  \,,
\end{multline}
and
\begin{multline}
  \label{fxcDave}
  {\cal Z}^\text{ph} (\xi)= \frac{1}{\tanh\left(\frac{\beta}{2}\xi\right)}\int_{-\mu}^{\infty} d\xi'
  \int\!\!\D\Omega\; \alpha^2F(\Omega) \\  \times
  \left[J(\xi,\xi',\Omega) + J(\xi,-\xi',\Omega)\right]
  \,,
\end{multline}
with  $I$ and $J$ given by Eqs.~(\ref{ikernel}) and (\ref{jkernel}).
Finally, the Eliashberg spectral function is the electron-phonon coupling
constant averaged on the Fermi surface
\begin{multline}
\label{a2F}
  \alpha^2F(\Omega) = \frac{1}{N(0)} \sum_{\nk,\nkp}\sum_{\lambda,\bq} 
  \left| g_{\lambda,\bq}^{\nk,\nkp} \right|^2 \\
  \times \delta(\xi_\nk) \delta(\xi_\nkp)\delta(\Omega - \Omega_{\lambda,\bq})
  \,.
\end{multline}
Note that in Eq. (\ref{a2F}) (and, consequently, in Eq.~(\ref{fxcCave}))  we replaced the 
density of states $N(\xi)$ by its value at the Fermi energy $N(0)$. 
This procedure is well justified, because it only requires that for each given band the
couplings do not change much on an energy scale of the order of the debye frequency, 
i.e., meV. The energy dependence of the density of states, on the other hand, is kept
in Eqs. (\ref{eq:gap_E}) and (\ref{eq:Z-ave}), as $N(\xi)$ can vary 
even on this small energy scale (e.g., in transition metals where the Fermi energy is
at the edge of large peaks in the density of states).

In order to evaluate the Coulomb terms, we further approximate the Kohn-Sham eigenvalues
by a free-electron (FE) parabolic dispersion $\varepsilon_{\bk}=k^2/2$. This approximation is
well justified for simple metals, but it is expected to fail for more complicated systems.
Within this approximation the energy-average of the Thomas-Fermi screened Coulomb 
interaction, given in Eq.~(\ref{melocal}), reads:
\be
  \label{eq:vTF_e}
  {\cal K}^{\rm TF-FE}(\xi,\xi') = \frac{\pi}{k k'}
  \log \left[\frac{\left(k+k'\right)^2+k^2_{\rm TF}}
    {\left(k-k'\right)^2+k^2_{\rm TF}}\right]
  \,,
\ee
with $k=\sqrt{2(\xi-\mu)}$ and $k'=\sqrt{2(\xi'-\mu)}$.

\subsection{Hybrid functionals}

It is possible to obtain a hybrid approach where the averaged functionals are used in the
$\bk$-dependent gap-equation~(\ref{eq:gap}). The averaged phononic terms (\ref{fxcCave}) and
(\ref{fxcDave}) can be used in ~(\ref{eq:gap}), just by replacing the energies $\xi$ and $\xi'$ 
by the energy eigenvalues of the real material $\xik$ and $\xikp$. 
\begin{eqnarray}
{\cal K}^\text{ph}_{\nk,\nkp} &=& {\cal K}^\text{ph}(\xi_{\nk},\xi_{\nkp}) \,, \\
{\cal Z}^\text{ph}_{\nk} &=& {\cal Z}^\text{ph}(\xi_{\nk}) \,.
\end{eqnarray}
For the electronic terms
we use an approach that goes along the lines of Sham and Kohn\cite{skmap} (SK), as
detailed in \PI. The basic idea is again to replace the free-electron bands by the real bands
of the material, and furthermore to adjust the chemical potentials of the two systems.
The functional reads
\begin{multline}
  \label{avecimpr}
  {\cal K}^\text{TF-SK}_{\nk,\nkp}=\frac{\pi}{2\sqrt{\eta_\nk \eta_\nkp}} \\
  \times \log\left(\frac{\eta_\nk+\eta_\nkp+2\sqrt{\eta_{\nk}\eta_\nkp}+k^2_\text{TF}/2}
  {\eta_\nk+\eta_\nkp-2\sqrt{\eta_\nk \eta_\nkp}+k^2_\text{TF}/2}\right)
\end{multline}
In this expression we defined $\eta_{\nk} = \xi_{\nk}+\frac{1}{2}k_\text{F}^2$, where $k_\text{F}$
is the Fermi wave-vector of a homogeneous electron gas with (constant) density equal to the
average density of the material.

\section{Computational Details}
\label{sec:compsec}

Obtaining superconducting properties through the solution of the gap equation can be viewed
as post-processing results of standard electronic structure calculations. In this work we
used electronic band structures obtained from full-potential linearized augmented 
plane wave~\cite{flapw} calculations; the same method is also used to compute the Coulomb 
potential matrix elements~(\ref{melocal}), along the lines described in Ref.~\onlinecite{matrixelem}.
On the other hand, the Eliashberg function can be obtained from linear response calculations~\cite{LRT}. 
(Note that the Eliashberg functions can also be extracted from experiment. While the
use of such experimental $\alpha^2F$'s renders the theory semi-phenomenological it
often gives useful insights.)

\begin{figure*}
  \includegraphics[width=0.85\columnwidth,clip]{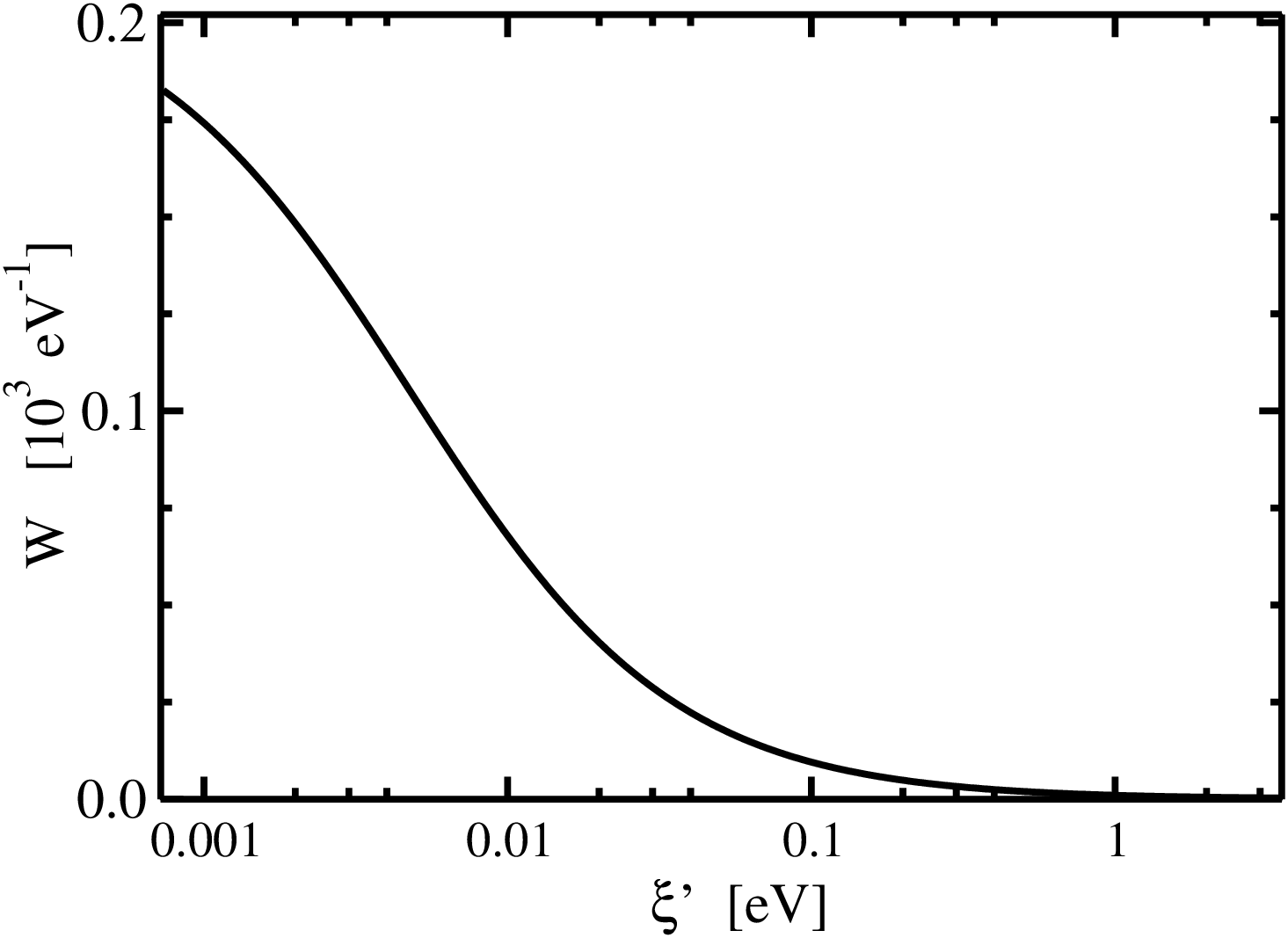}
  \qquad
  \includegraphics[width=0.85\columnwidth,clip]{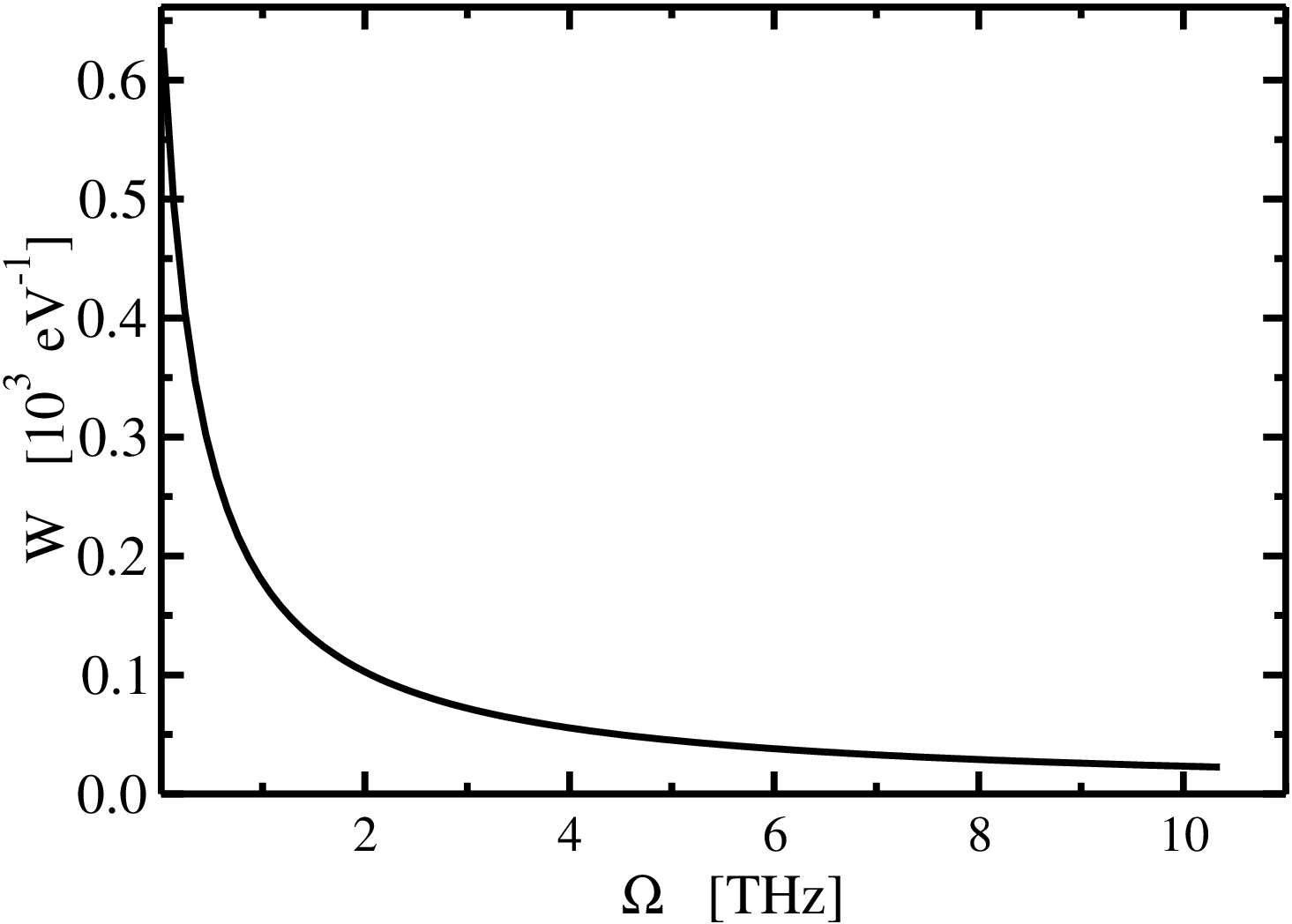}
\caption{Universal kernel $W(\xi,\xi',\Omega)$ as a function of $\xi'$ and $\Omega$.
Left panel: universal kernel $W(\xi\approx 0,\xi',\Omega=1\text{\,THz})$ as a function of $\xi'$.
Right panel: $W(\xi=\xi'\approx 0,\Omega)$ as a function of $\Omega$. Both panels were
obtained for $T=0.01$\,K. 
\label{kern}}
\end{figure*}
We developed two completely independent codes: the first solves the averaged 
gap equation~(\ref{eq:gap_E}) in energy space; the second solves the $\bk$-resolved
gap equation~(\ref{eq:gap}). The two schemes were compared whenever possible, and give
similar results for the simple metals studied here (see the discussion at the 
end of this section). The solution of the
averaged gap equation is a quite simple numerical task. The energy is discretized in
a logarithmic mesh in order to increase the accuracy close to the Fermi level and the
resulting equation is iterated until convergence. This code was used within the
TF-FE scheme. (For a summary of the different schemes, please refer to Table~\ref{tab:approx}.)
On the other hand, to use the TF-ME and the TF-SK functionals we need to
solve the $\bk$-resolved equation. This is a difficult numerical task mainly because:
i) we need to describe the energy bands over a large energy window and, at the same 
time, ii) we require a very good resolution in a small energy window around the Fermi
energy. Requirement i) comes from the large energy scales involved in the
Thomas-Fermi screened Coulomb potential (typically of the order of the Fermi energy).
This is not an artifact of the static nature of the Thomas-Fermi screening. In fact,
even a more realistic dynamically screened  potential would need a very large energy 
range. This point  will be further discussed in the following section. 
To illustrate requirement ii) we plot, in Fig.~(\ref{kern}), the kernel of Eq.~(\ref{fxcCave}),
\be
  W(\xi,\xi',\Omega)= I(\xi,\xi',\Omega)- I(\xi,-\xi',\Omega)
  \,.
\ee
The left panel of Fig.~\ref{kern} depicts the variation of $W$ with energy. This plot
was obtained at very low temperature $T=0.01$\,K, by fixing $\xi'$ very close to the 
Fermi surface $\xi\approx 0$. (We did not use the values $T=0$ and
$\xi=0$ to avoid numerical problems.) Considering that the kernel decreases to about to about 20\% of 
its value within roughly 10 meV, it is clear that we need to sample very carefully 
this energy range. This is particularly important for materials, like niobium 
and tantalum, where the Fermi energy is at the edge of a peak in the density of states.
In the right panel we plot the dependence of $W$ on the frequency
$\Omega$, for $\xi$ and $\xi'$ very close to the Fermi surface. 
The kernel $W$ acts as a weight for $\alpha^2F(\Omega)$ (see, e.g., 
Eq.~(\ref{fxcCave})), and enhances the contribution of the lower-frequency phonons.

In order to fulfill the requirements i) and ii) we developed the following numerical framework:
Starting from ab-initio bands calculated over a few hundred $\bk$-points in the irreducible wedge 
of the Brillouin zone (BZ), we compute a good spline fit of $\xik$ over a Fourier series, according to 
the scheme of Koelling and Wood~\cite{koelling}. Using this fit, we then obtain the energies $\xi$ over a 
very large set of random $\bk$-points, distributed according to a Metropolis algorithm. This
algorithm was devised to accumulate a large number of $\bk$-points in the first few meV's around the Fermi
surface. A good convergence can be reached by using about 15000--20000 independent points for each band 
crossing the Fermi surface, while a reasonable description of the remaining bands is obtained with about 1000 
independent points per band. The \DFT\ scheme is the slowest to converge, as it involves the integration
of a function which is peaked around $|\bk'-\bk | \to 0$ (the integrand would diverge if we used the bare 
Coulomb potential instead of the Thomas-Fermi screened interaction).

The function $\Delta_\nk$ only needs to be defined in the irreducible wedge of the Brillouin zone. 
On the other hand, when using the TF-ME scheme, the $\bk'$ summation has to be performed over the 
whole zone. This can be seen from the expression of the Coulomb matrix elements, Eq.~(\ref{melocal}): 
By setting $\bk\neq 0$ the symmetry of the $\bk'$ summation is broken (only the operations of the 
little  group of $\bk$ can be retained). This is a well-known situation also in the framework of electronic 
self-energy calculations. The case of the hybrid functionals is simpler. As these functionals depend
on $\bk$ and $\bk'$ though $\xi_\nk$ and $\xi_\nkp$, and as the eigenvalues $\xik$ are totally
symmetric with respect to the crystal point group, the summation over $\bk'$ can be limited
to the irreducible wedge of the Brillouin zone. This formalism can be easily generalized to the 
case of non $s$-wave pairing, by imposing that the gap transforms according to a given irreducible 
representation of the crystal point group. 

As an initial guess for the gap function $\Delta_\nk$ we use a step function.
By using a Broyden scheme\cite{broyd} to mix $\Delta_\nk$, we obtain convergence
in a mere 10--15 self-consistent iterations. The converged result at a given temperature can be used as a 
starting point for the next temperature. Clearly, this procedure reduces the total number of 
iterations required.

The matrix elements of the screened Coulomb potential are obtained with the full-potential linearized augmented 
plane wave method, as explained in Ref.~\onlinecite{matrixelem}. They are calculated for all bands within
15--30\,eV from the Fermi surface (typically 14 valence and conduction bands). It is
unfeasible to obtain the matrix elements for the whole set of random $\bk$-points 
used in the solution of the gap equation. Thus, we calculate the matrix elements for a set 
of $\bk$ and $\bk'$ belonging to a regular mesh. The values at the random points $\bk$ and $\bk'$ 
are then obtained by mapping $\bk$ and $\bk'$ to the corresponding hypercube of the regular mesh.
Test calculations have shown that already a $6\times 6\times 6$ mesh gives acceptably small residual 
errors (1-2 \% of the gap at $E_F$).
A finer mesh is however needed to quantitatively estimate the spread of the gap for each 
given energy, which is material dependent and 
mostly due to the physical $\bk$ and $\bk'$ dependence of the matrix elements.  

\begin{figure}
\begin{center}
\includegraphics[width=0.85\columnwidth,clip]{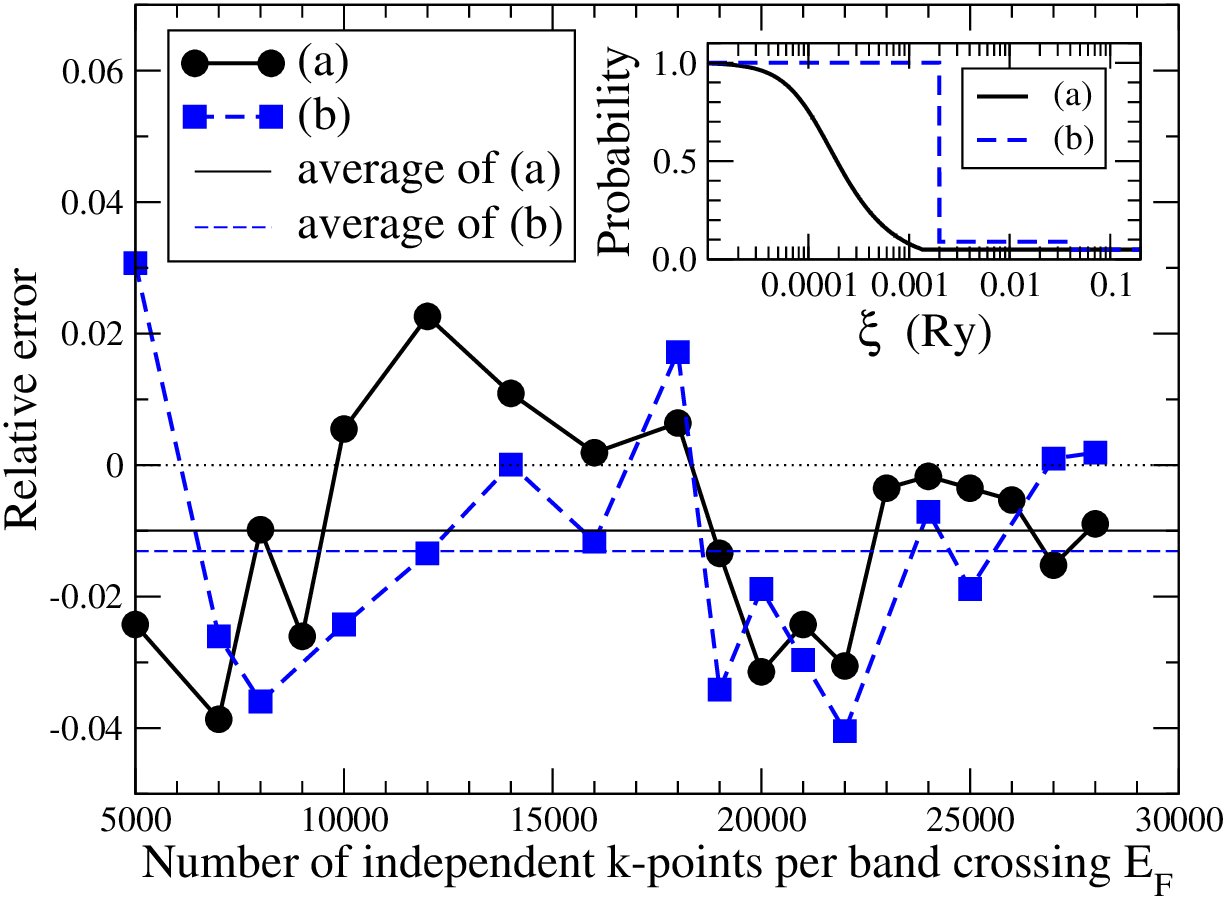}
\end{center}
\caption{\label{fig:conv_kpt} (Color online)
Relative difference between the $\bk$-resolved  and the energy-integrated results for Nb,
as a function of the number of independent $\bk$-points used for each band crossing
$E_F$. The circles and squares are obtained, respectively, with probability distributions
equal to $max\{\chi(\xi,T=0),0.05\}$  and to a stepwise function having widths of 2 and 40
mRy, as shown in the insert. Thin lines represent the arithmetic averages of the k-resolved
data. 
\label{convk}}
\end{figure}

In Fig.~\ref{convk} we show a detailed analysis of the convergence of our method for Nb,
using the TF-SK approach.  In this figure we plot the relative differences
between the  {\bf k}-resolved approach and the results of the
energy-only scheme, which can be considered to be numerically convergent
(because of its use of a logarithmic energy mesh), but itself depending on the fine details
of the input density of states. This is particularly true in the case of
Nb, where $E_F$ sits in a rapidly varying part of the DOS.
The results are shown as a function of the number of independent k-points
used for each band crossing the Fermi level, and circles and squares
represent the results obtained using two different distribution functions used
to sample the Brillouin zone (as detailed in the figure caption).
We can see  the capability of our procedure to converge towards the
(computationally completely independent) energy-only result, with  a very small
numerical error. The two rather different samplings  lead to
almost identical averages, with a correct, gaussian-like, distribution of results
around the average.
We notice that bands away from the $E_F$  are sampled uniformly in {\bf k}-space,
as it should since at these energies only the Coulomb term remains, with a weak energy
dependence.

For simplicity, we used in all our calculations the averaged or hybrid phononic
functionals. In this case, the electron-phonon interaction enters the calculation through
the Eliashberg function $\alpha^2F(\Omega)$. This corresponds to neglecting the anisotropy
of the electron-phonon coupling, which should be a good approximation for the particular
systems studied in the following.

We quantify the precision of our results to be around 5\%. This value includes the
uncertainty associated with the Metropolis procedure. We emphasize that this is a 
non-trivial numerical achievement, as superconducting properties depend exponentially
on the electron-phonon coupling and on the Coulomb interaction.
Particularly difficult is to obtain numerically stable 
results for the small gap materials, where superconductivity stems from an almost complete 
cancellation of large and opposite terms.
 
\section{Results and discussion}
\label{sec:res}

\begin{table*}
\begin{tabular}{cccc}
Method & Gap-equation & Phononic terms & Coulomb term \\ \hline \\[-0.2cm]
TF-ME  & $\bk$-dependent
       & Hybrid: averaged with real bands
       & Matrix elements \\
       & Eq.~(\ref{eq:gap}) & Eqs.~(\ref{fxcCave}) and (\ref{fxcDave}) & Eq.~(\ref{melocal})
\\[0.1cm]
TF-SK  & $\bk$-dependent
       & Hybrid: averaged with real bands
       & Hybrid: averaged with real bands \\
       & Eq.~(\ref{eq:gap}) & Eqs.~(\ref{fxcCave}) and (\ref{fxcDave}) & Eq.~(\ref{avecimpr})
\\[0.1cm]
TF-FE  & Energy averaged 
       & Averaged
       & Averaged \\
       & Eq.~(\ref{eq:gap_E}) & Eqs.~(\ref{fxcCave}) and (\ref{fxcDave}) & Eq.~(\ref{eq:vTF_e})
\end{tabular}
\caption{Summary of the different methods used in our calculations.
\label{tab:approx}}
\end{table*}

In the following we will present results obtained for the simple metals Al, Mo, Ta, Nb, and Pb.
Note that this group of materials includes both weak coupling (Al, Mo, and Ta) and strong coupling
(Nb and Pb) superconductors. We solved the gap equation within three different approaches,
that we label \LDAzero, \LDAone, and \DFT\ (see Table~\ref{tab:approx}).

\begin{figure}
\begin{center}
\includegraphics[width=0.85\columnwidth,clip]{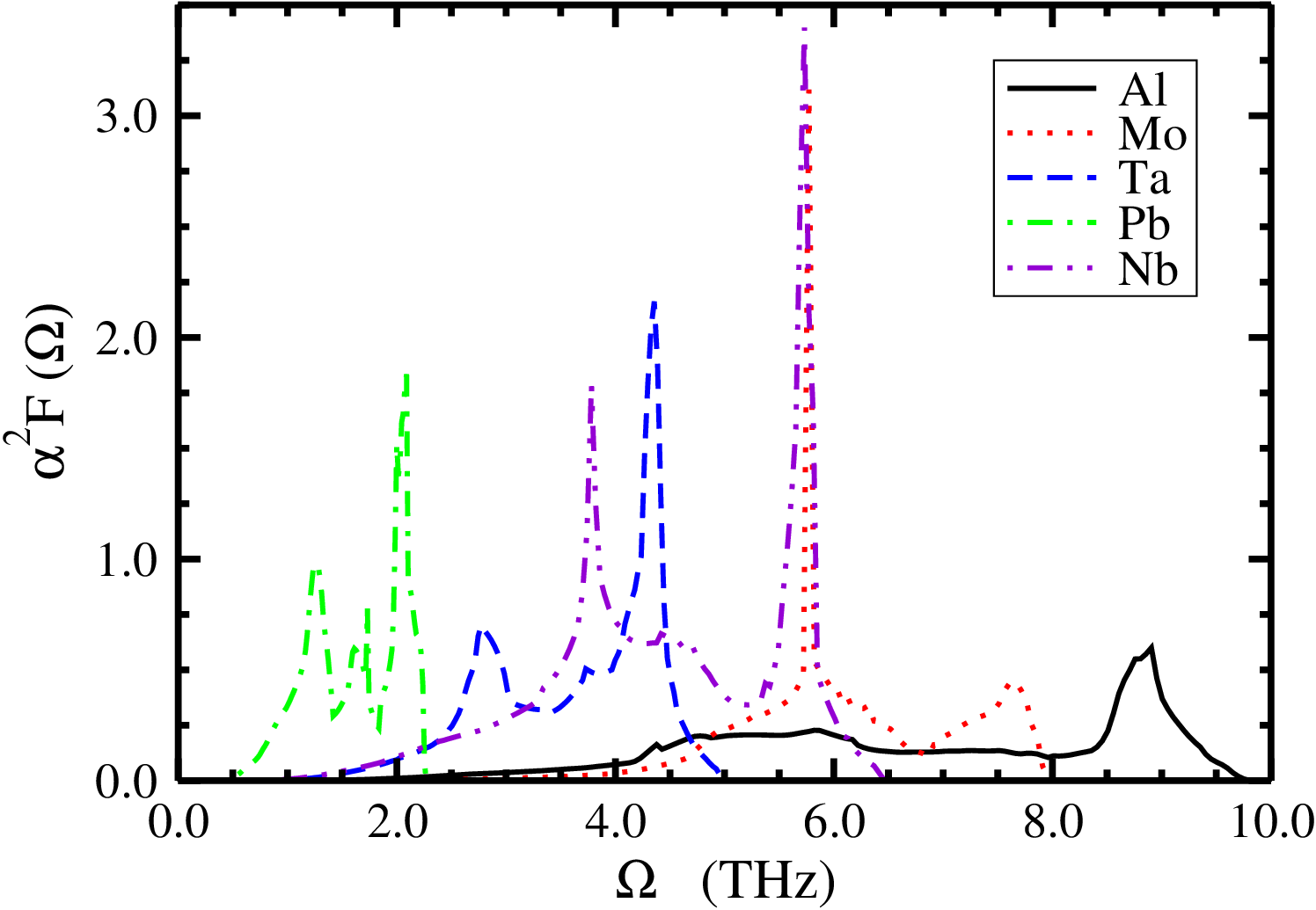}
\end{center}
\caption{(Color online) 
The Eliashberg function $\alpha^2F(\Omega)$ for the simple metals used in our 
calculations. The curves are taken from Ref.~\onlinecite{savrasov}.
\label{a2fig}}
\end{figure}
In Fig.~\ref{a2fig} we show the Eliashberg functions used in our calculations.
All these curves were obtained by Savrasov\cite{savrasov} using linear response theory.
The five materials cover a broad frequency range, from 1 to 10\,THz. The $\alpha^2F(\Omega)$
function for Pb is located at much lower frequencies than for the other materials, due
to the heavy nuclear mass of Pb. For this reason, Pb has the largest total electron-phonon 
coupling constant $\lambda$ of the materials studied 
($\lambda = 2 \int\!\!\D\Omega\; \alpha^2F(\Omega)/\Omega$). Nevertheless, Nb becomes superconducting
at higher temperatures than Pb. This can be easily understood on the basis of McMillan's
formula:~\cite{mcmillan} Besides the well-know exponential dependence on the electron-phonon coupling constant,
the transition temperature increases linearly with the average phonon frequency.
Lowering the phonon frequencies has therefore both a positive and a negative effect
on $T_\text{c}$. An accurate description of the superconducting state must take into account
both these effects.

\begin{figure*}
\begin{center}
  \includegraphics[width=0.85\columnwidth,clip]{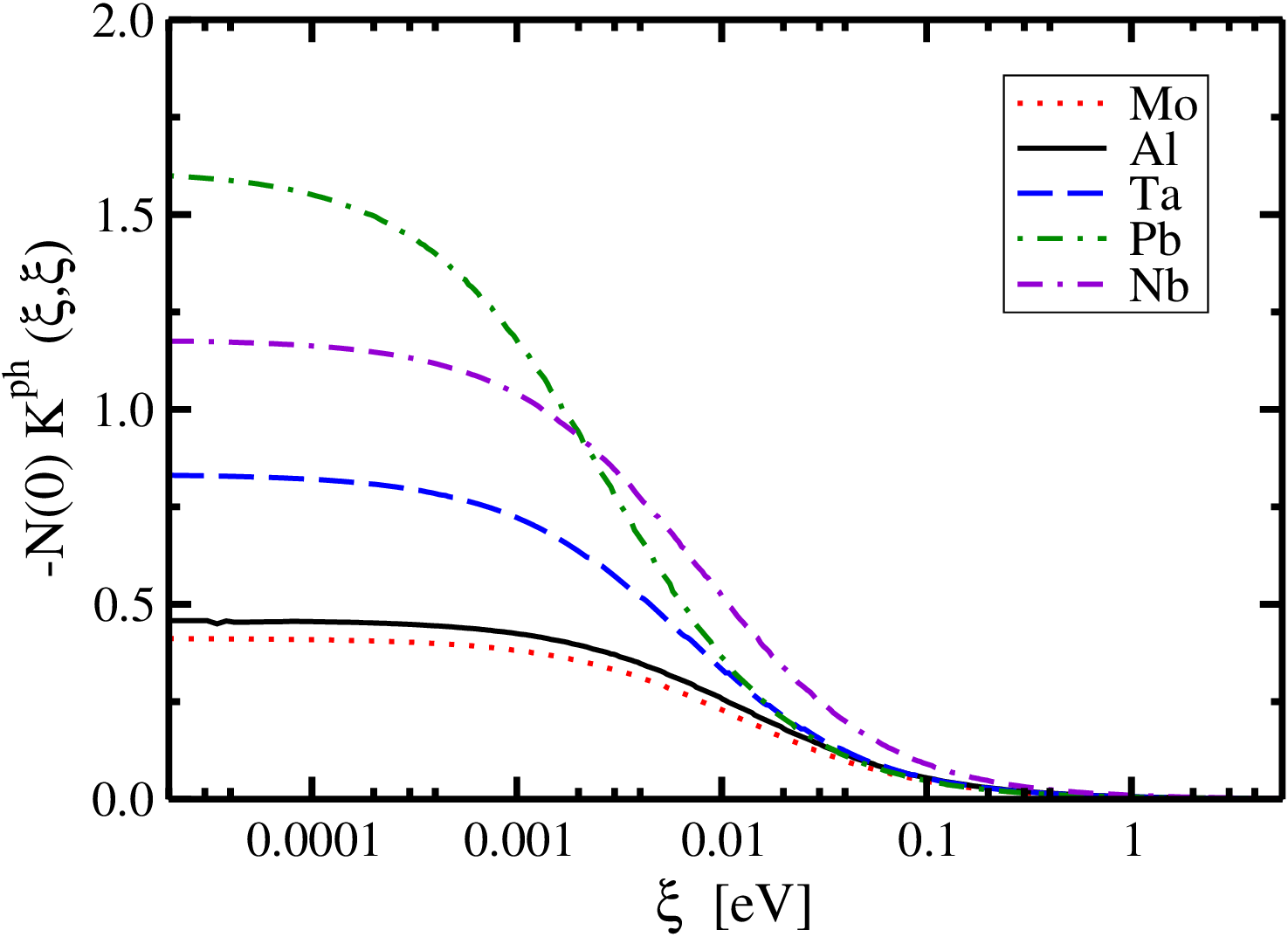}
  \qquad
  \includegraphics[width=0.85\columnwidth,clip]{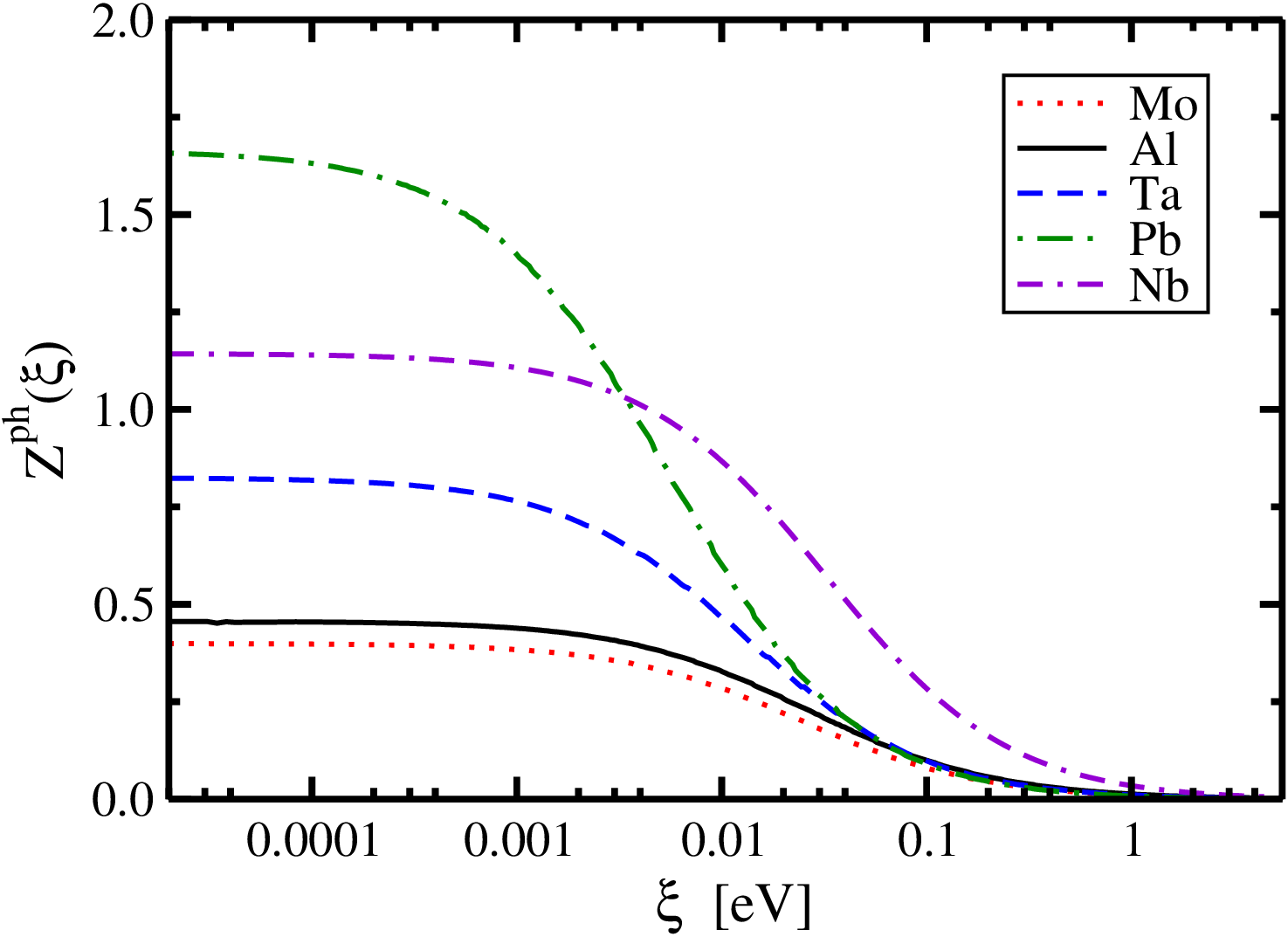}
\end{center}
\caption{(Color online) 
Diagonal of $-N(0){\cal K}^\text{ph}(\xi,\xi')$ (left panel) and ${\cal Z}^\text{ph}(\xi)$
at $T=0.01$\,K for the simple metals studied.
\label{cdterm}}
\end{figure*}
Before examining our results, we depict in Fig.~\ref{cdterm} the behavior of the phononic functionals,
$-N(0){\cal K}^\text{ph}(\xi,\xi')$ and ${\cal Z}^\text{ph}(\xi)$. Both terms turn out to be very
peaked at the Fermi energy, and are non-zero only in a small region around it
(of the order of the Debye energy). Furthermore, the values of the functionals at the Fermi 
energy read $N(0){\cal K}^\text{ph}(0,0)=-\lambda$, and ${\cal Z}^\text{ph}(0)=\lambda$. The
${\cal K}^\text{ph}$ term is negative, reflecting the electron-phonon attraction responsible for
the superconducting state. On the other hand, the term ${\cal Z}^\text{ph}$ is positive and
tends to decrease the superconducting gap and therefore $T_\text{c}$.

\begin{figure*}
\begin{center}
  \includegraphics[width=0.85\columnwidth,clip]{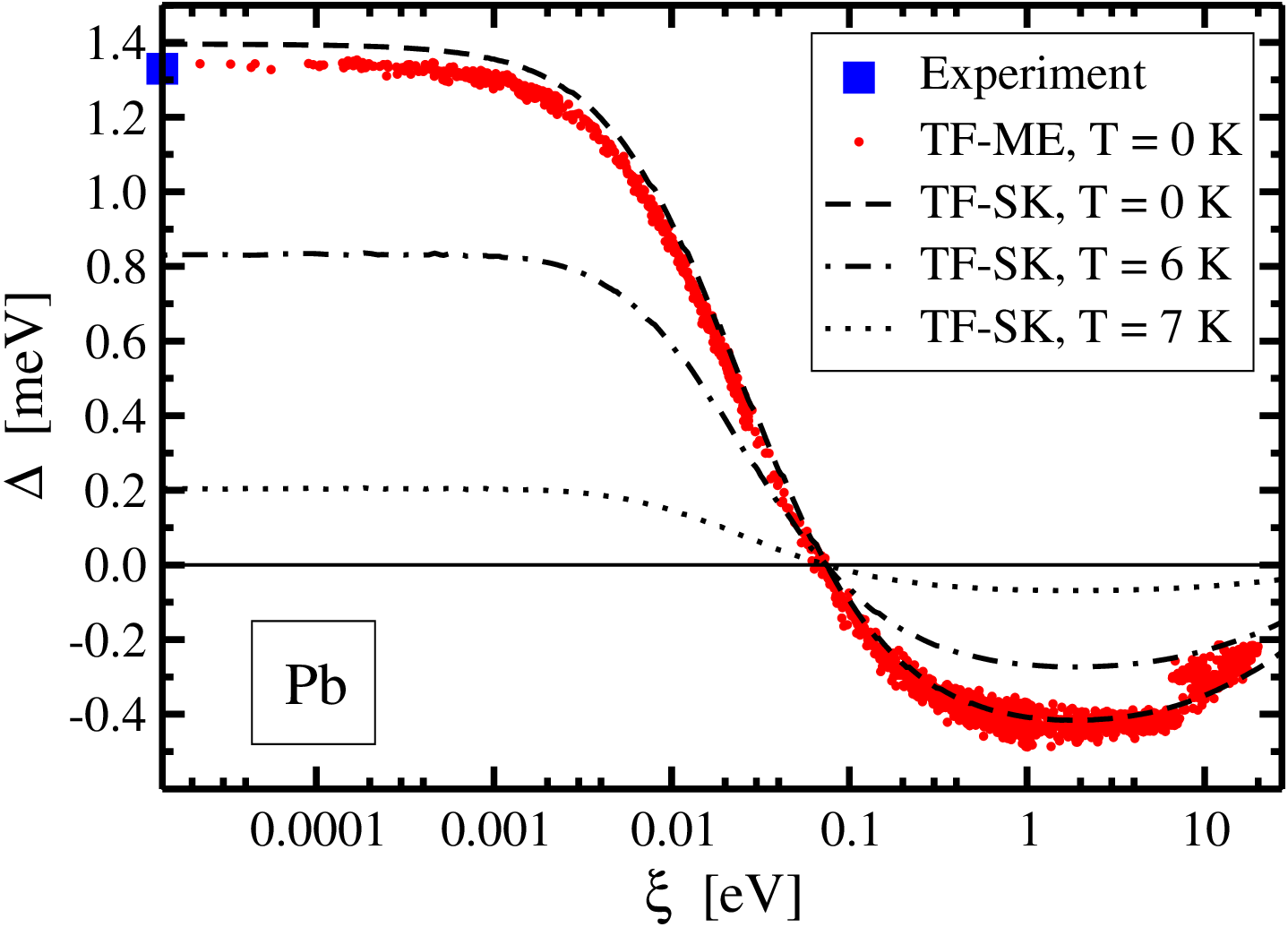}
  \qquad
  \includegraphics[width=0.85\columnwidth,clip]{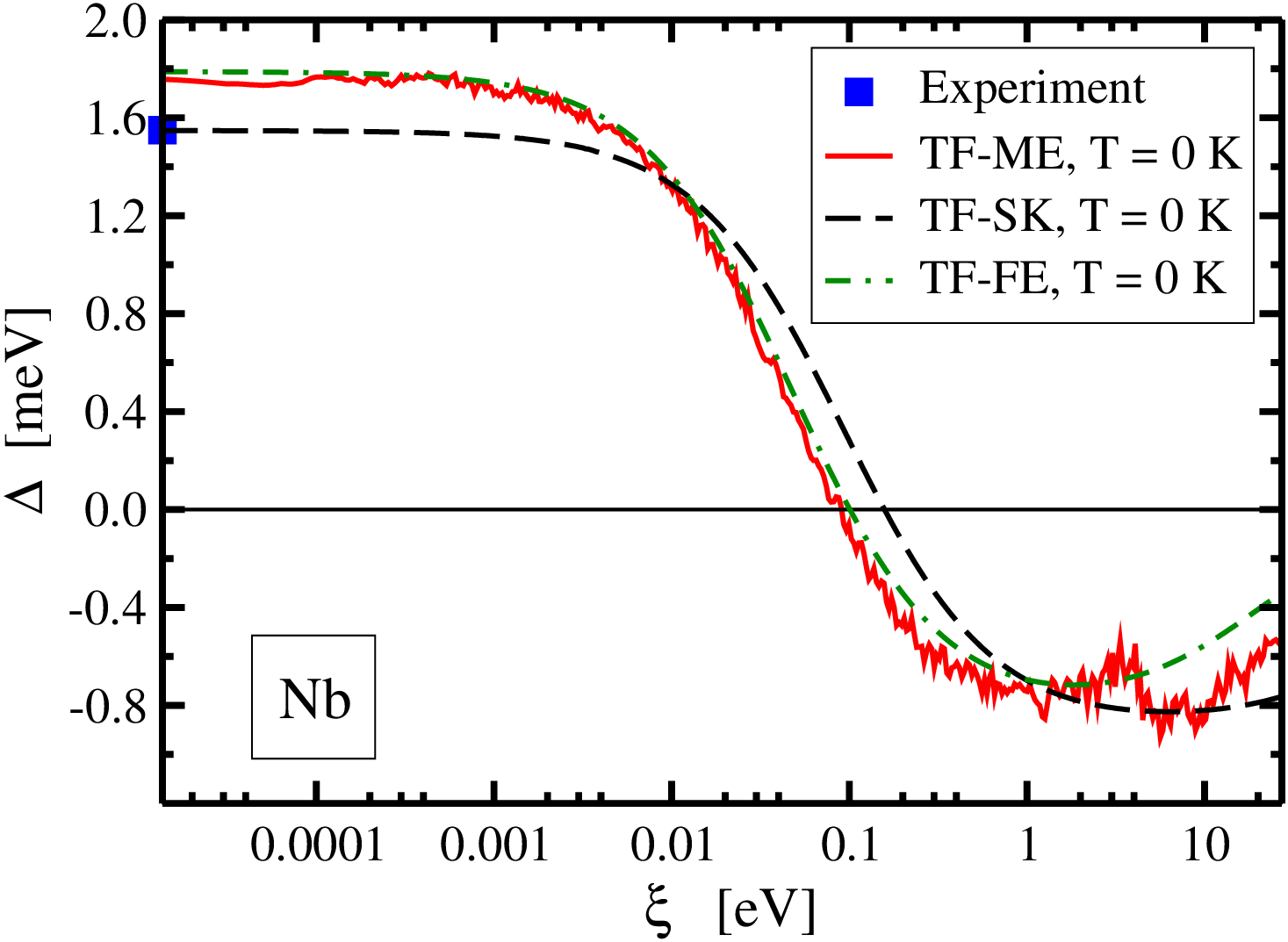}
\end{center}
\caption{(Color online) 
The function $\Delta(\xik,T)$ for lead (left panel) and niobium (right panel).
\label{pb_nb_gap_en}}
\end{figure*}
In Fig.~\ref{pb_nb_gap_en} we show the superconducting gap $\Delta$ calculated using different
approaches for Pb and Nb. For Pb we also show curves calculated at different temperatures.
It is important to note that for $T>T_\text{c}$ the self-consistent calculation correctly 
converges to zero even if the starting gap function is different from zero. The curves 
labeled \DFT\ show the detailed sampling of the whole energy range given by our $\bk$-point mesh, down to
a few $\mu$eV. Within the TF-ME approach, the spread in the values of the gap function of Pb 
at energies near $E_F$ 
is quite small (around 2\%), which can be understood as a consequence of a rather isotropic 
Fermi surface. This spread is larger for the higher energy states because of the larger spread of 
orbital character in the electronic states. 
The gap function for both materials exhibits a very similar shape, 
with a node at about 0.4\,meV, and a negative tail extending to high energy. This shape is
basically independent of the temperature, and is a general feature found in all our calculations.
Due to the presence of Coulomb repulsion, this shape is a necessary condition to obtain a superconducting
solution. It is well known that the structure of the gap equation is such that the repulsive 
Coulomb interaction between two Cooper pairs (at $\{\bk\uparrow,-\bk\downarrow\}$ and 
$\{\bk'\uparrow,-\bk'\downarrow\}$) gives a constructive contribution to the gap if the values of $\Delta_{\nk}$ 
and $\Delta_{\nkp}$  have opposite signs. This condition is realized when, e.g., $\xik$ is small and 
$\xi_{\nkp}$ is large. Similar arguments are the basis of the classical calculations of Ref.~\onlinecite{morel}, 
and result in the definition of the renormalized Coulomb parameter $\mu^*$.

The three different schemes (\LDAzero, \LDAone, and \DFT) agree well among themselves for Pb, while
for Nb the \LDAone\ curve differs by 10\% from the \LDAone\ and \DFT\ results. This difference
can be explained by looking at the band structure of the materials. In Pb, the valence and 
conduction bands are basically due to $s$ and $p$ orbitals, for which the averaged and hybrid schemes
work quite well. The same result is found for Al, with an even larger similarity between \LDAzero, 
\LDAone\ and \DFT\ results. All this is easily understood from the fact that the three 
schemes \LDAzero, \LDAone, and \DFT\ become identical in the limit of the uniform gas. Hence one
would expect them to yield similar results for the delocalized s-p orbitals. On the other hand, 
the bands of Nb exhibit a strong $d$-character, which 
leads to the difference between the methods. A similar behavior is found for Ta. This 
trend is also confirmed by the  value of the gap function for the low lying semi-core $d$ 
states of Pb (not shown in the figure) which is very different in the \LDAone, and 
\DFT\ calculations (it is much smaller for \DFT). This is a clear consequence of the 
localized nature of these states, implying a very large repulsive 
self-term for those bands. In principle, the \DFT\ approach should be the most precise.

Note that our approximate functionals use the Thomas-Fermi model for screening. We have 
compared the matrix elements of the Thomas-Fermi-screened potential with those coming
from a complete (static) RPA and found very close agreement between the two approaches. 
It is very important in this context to define the Thomas-Fermi wave vector $k_{TF}$
in terms of the density of states at the Fermi level, $N(0)$. As the latter is obtained 
from a full-scale Kohn-Sham calculation, $k_{TF}$ is treated here implicitly as a rather
complicated density functional. 

Furthermore, we neglect the off-diagonal elements of the dielectric matrix. This is  not necessarily 
a good approximation for transition metal compounds. However, it was shown in the literature
\cite{Chang95,Chang96} that the changes produced by these local field effects are largely 
cancelled by the additional inclusion of exchange-correlation effects (generated by the 
exchange-correlation kernel \cite{GrossKohn85} of time-dependent density functional theory 
\cite{RungeGross84}).
We believe that our good results, obtained by neglecting both local-field and exchange-correlation
effects, can at least partly be explained by this cancellation. Further help comes from
the fact that superconductivity describes correlations on the scale
of the coherence length, which is usually much larger than the dimensions of the unit cell. 
On this scale, local field effects are certainly less important.

\begin{figure*}
\begin{center}
  \includegraphics[width=0.85\columnwidth,clip]{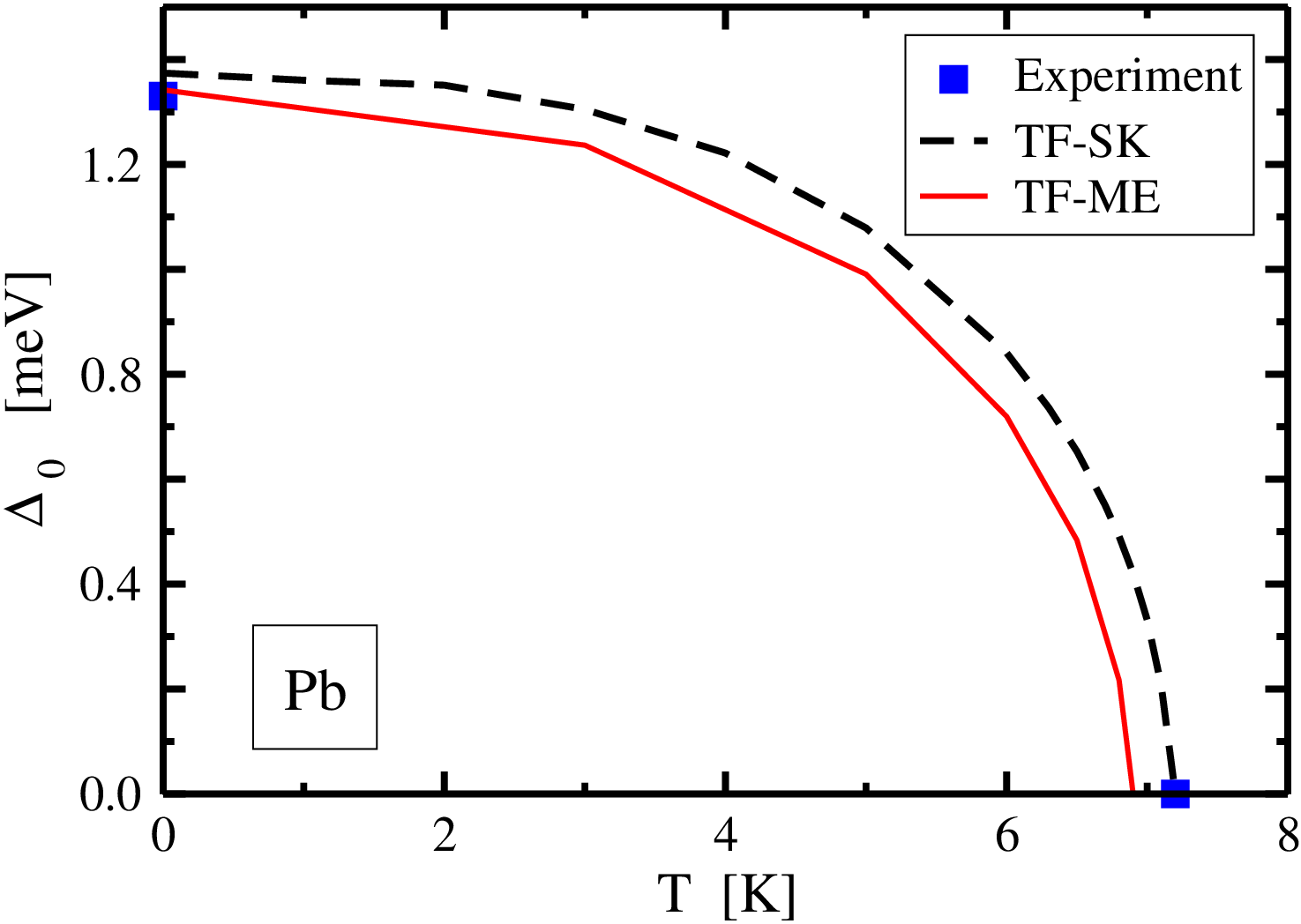}
  \qquad
  \includegraphics[width=0.85\columnwidth,clip]{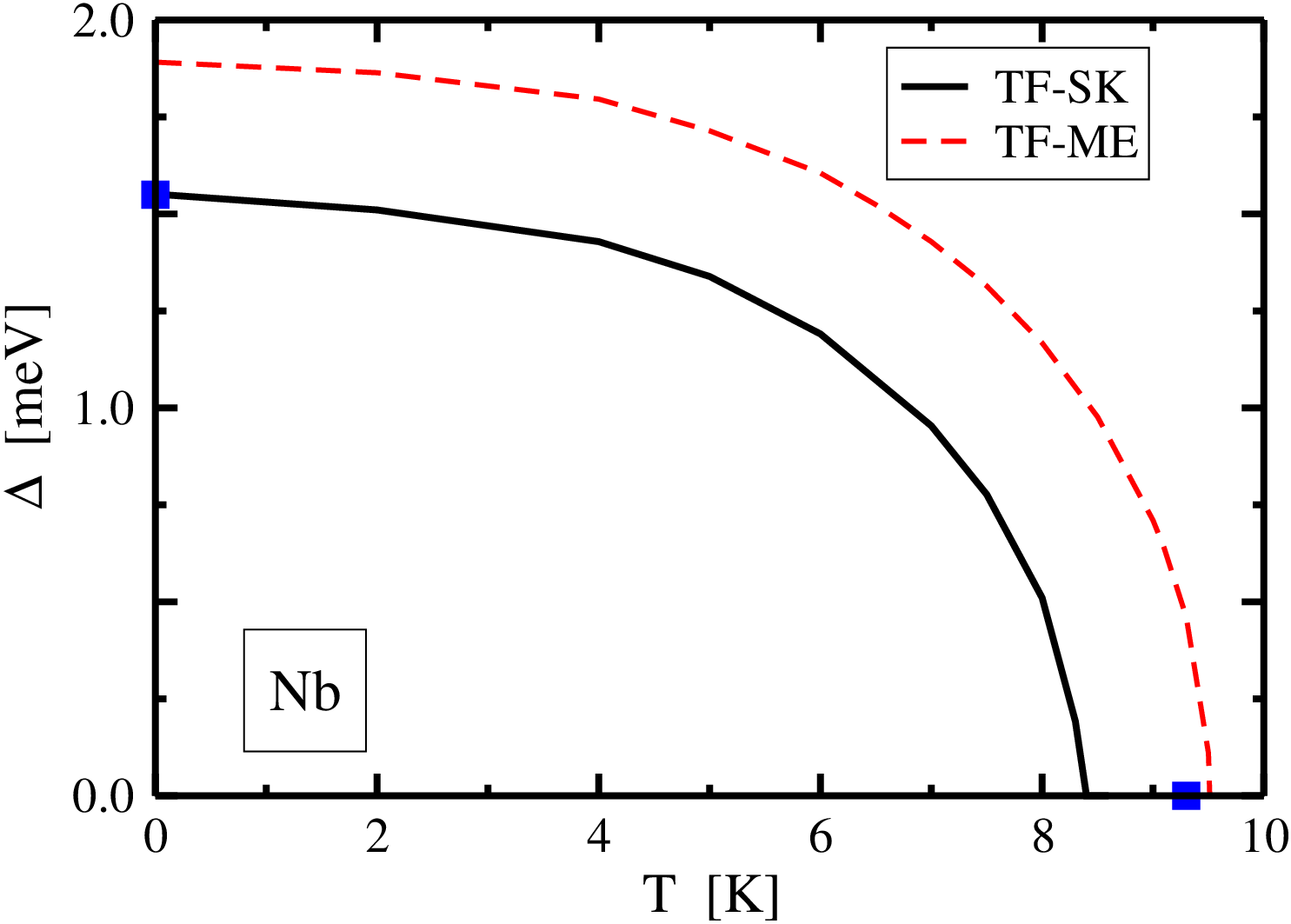}
\end{center}
\caption{(Color online)
The gap function at the Fermi surface $\Delta_0$ for Pb and Nb as a function of temperature.
  \label{Pb_Nb_T}}
\end{figure*}
The gap function at the Fermi energy is depicted in Fig.~\ref{Pb_Nb_T} as a function of
temperature for Pb and Nb. The curves show a BCS-like square root behavior close to
$T_\text{c}$, which we would expect for these simple superconductors. For Pb, both
$T_\text{c}$ and $\Delta_0$ agree quite well among themselves and with the experimental results.
For Nb, on the other hand, the TF-SK gap is roughly 15\% smaller than the TF-ME curve.
Curiously, the TF-SK gap at zero temperature is very close to the experimental value,
while the TF-ME yields a much better transition temperature. The difference between the
methods can be once more explained by the $d$-character of the bands in Nb close to the
Fermi energy.

\begin{table}
\begin{center}
\begin{tabular}{cccccc}
 &\multicolumn{5}{c}{$T_\text{c}$ [K]} \\
   & \DFT & \LDAone & \LDAzero &  exp    &  $\lambda$   
   \\ \hline
Mo & ---   & 0.33  &   0.54   &  0.92            & 0.42 \\
Al & 0.90  & 0.90  &   1.0   &  1.18             & 0.44 \\
Ta & 3.7   & 2.7  &   4.8   &  4.48             & 0.84 \\
Pb & 6.9   & 7.2  &   6.8   &  7.2              & 1.62 \\
Nb & 9.5   & 8.4  &   9.4   &  9.3              & 1.18
\end{tabular}

\begin{tabular}{cccccc}
 &\multicolumn{5}{c}{$\Delta_0$ [meV]} \\ 
   &  \DFT & \LDAone & \LDAzero &  exp  &  $\lambda$
   \\ \hline
Mo & ---   & 0.049 &  0.099   &  ----              & 0.42 \\
Al & 0.14  & 0.15  &  0.15    & 0.179              & 0.44 \\
Ta & 0.63  & 0.53  &  0.76    & 0.694              & 0.84\\
Pb & 1.34  & 1.40  &  1.31    & 1.33               & 1.62\\
Nb & 1.74  & 1.54  &  1.79    & 1.55               & 1.18
\end{tabular}
\caption{The critical temperature (upper panel) and the superconducting gap at Fermi level 
and $T=0.01$\,K (lower panel),   
  compared with experiment~\cite{ashcroft}. We show also the total electron-phonon coupling constant 
  $\lambda$ \cite{savrasov}.
  \label{tbresfr}}
\end{center}
\end{table}
\begin{figure*}
\begin{center}
  \includegraphics[width=0.75\columnwidth,clip]{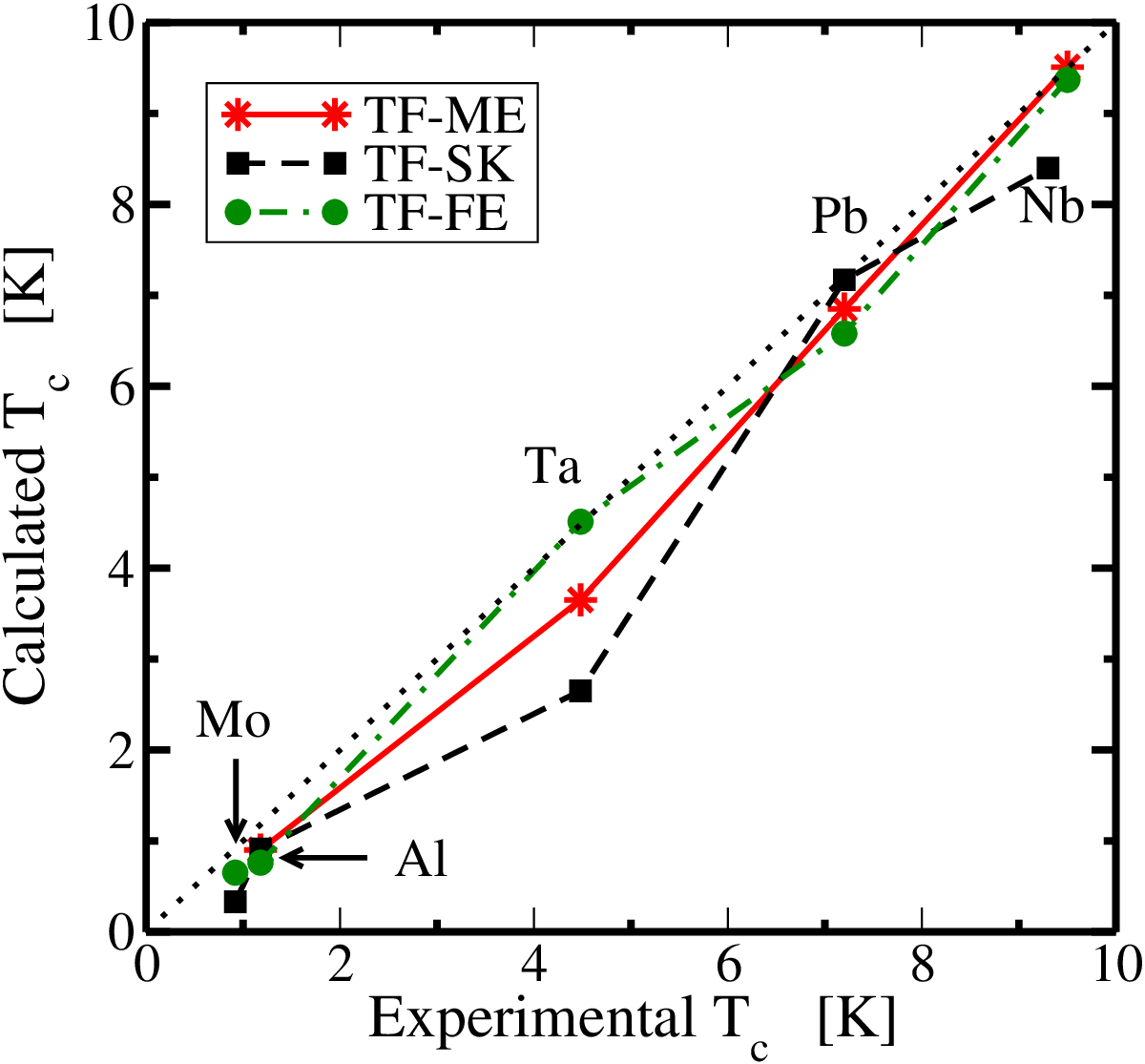}
  \qquad
  \includegraphics[width=0.75\columnwidth,clip]{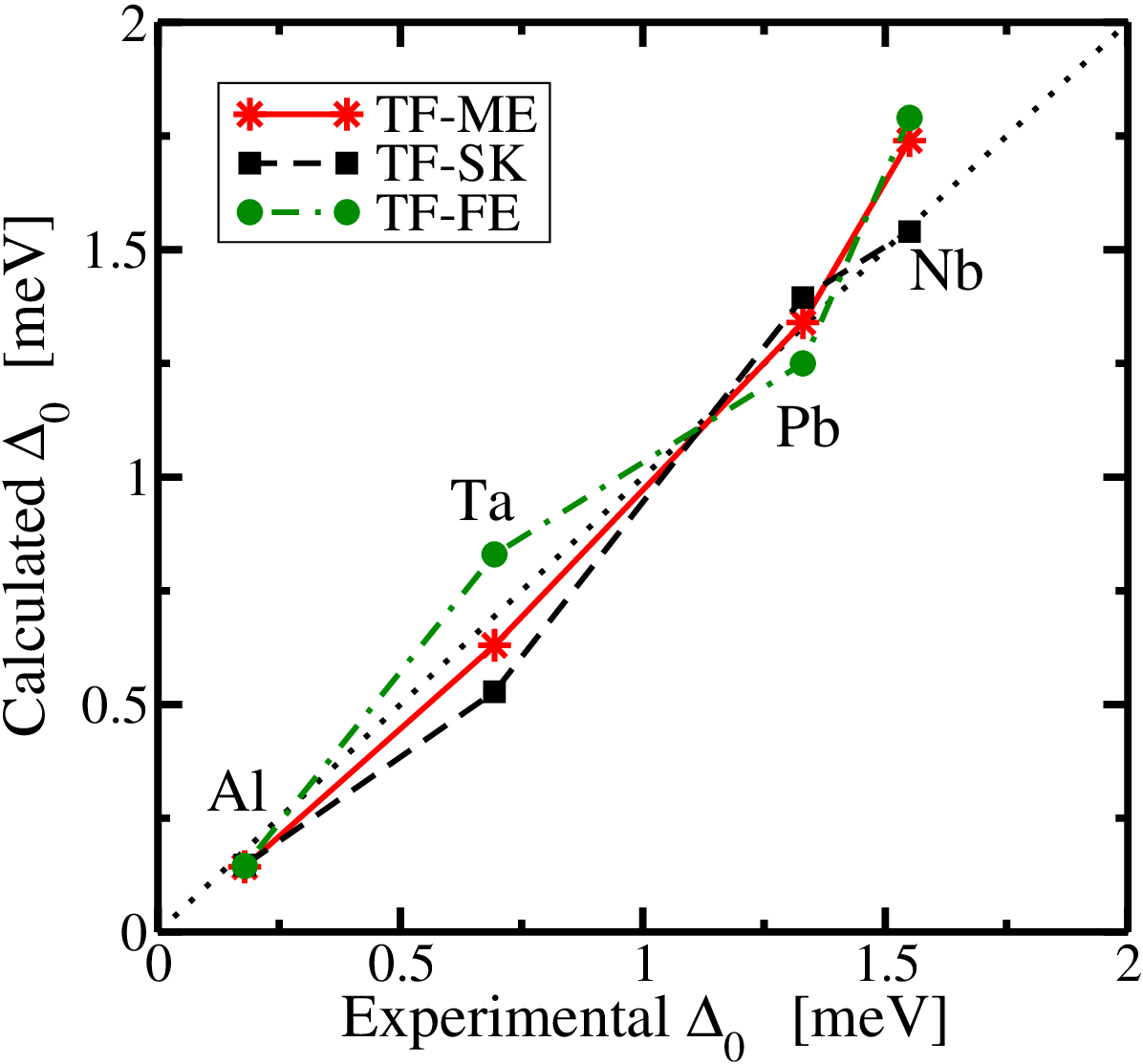}
\end{center}
\caption{(Color online)
The critical temperature and, superconducting gap at Fermi level and $T=0.01$\,K,   
  compared with experiment. The numerical values can be found in Table~\ref{tbresfr}.
  \label{figresfr}}
\end{figure*}
Our results for the superconducting transition temperatures and gap functions are summarized in
Table~\ref{tbresfr} and in Fig.~\ref{figresfr} for all materials studied. In the same table we also show the experimental
results, and the values of the electron-phonon coupling constant $\lambda$. Mo is a weak coupling
superconductor with a very small gap and very low transition temperature. In this case, the
different theoretical approaches lead to quite different results, and the overall agreement
with experiment is not very good. Surprisingly, it is the \LDAzero\ approach which gives the 
better results. However, many tests showed that the Mo results are very sensitive to the details 
of the density of states underlying our calculations.  This is quite normal for a material
with such a small gap resulting from a very fine balance between large terms.
In Al the electronic states have a strong free-electron character and the density of
states follows closely the free-electron DOS. It is not surprising, therefore, that all three
methods yield very similar results. These results are also in satisfactory agreement with experiment.
Ta shows a behavior similar to Nb, but the agreement with experiment is poorer. 
Moreover, the larger spread of values for Ta relative to Nb can be explained by the smaller values 
of $\Delta$ and $T_c$.  As in the case of Mo, the \LDAzero\ approach yields the best results.
This surprising result can be understood to some extent: The \LDAzero\ approach is fully consistent,
in the sense that all the quantities entering the method are averaged in similar ways; In the 
\LDAone\ approach, on the other hand,
certain quantities (but not all) are calculated at the average density. 
We emphasize, however, that the agreement 
of our results with experiment, without making use of any adjustable parameters, is unprecedented 
in the field.  
We note in passing that for Cu we could not find a non-trivial solution of the gap equation at
any temperature, which also in in agreement with the experimentally observed absence of 
superconductivity in Cu.

\begin{figure}
\begin{center}
  \includegraphics[width=0.85\columnwidth,clip]{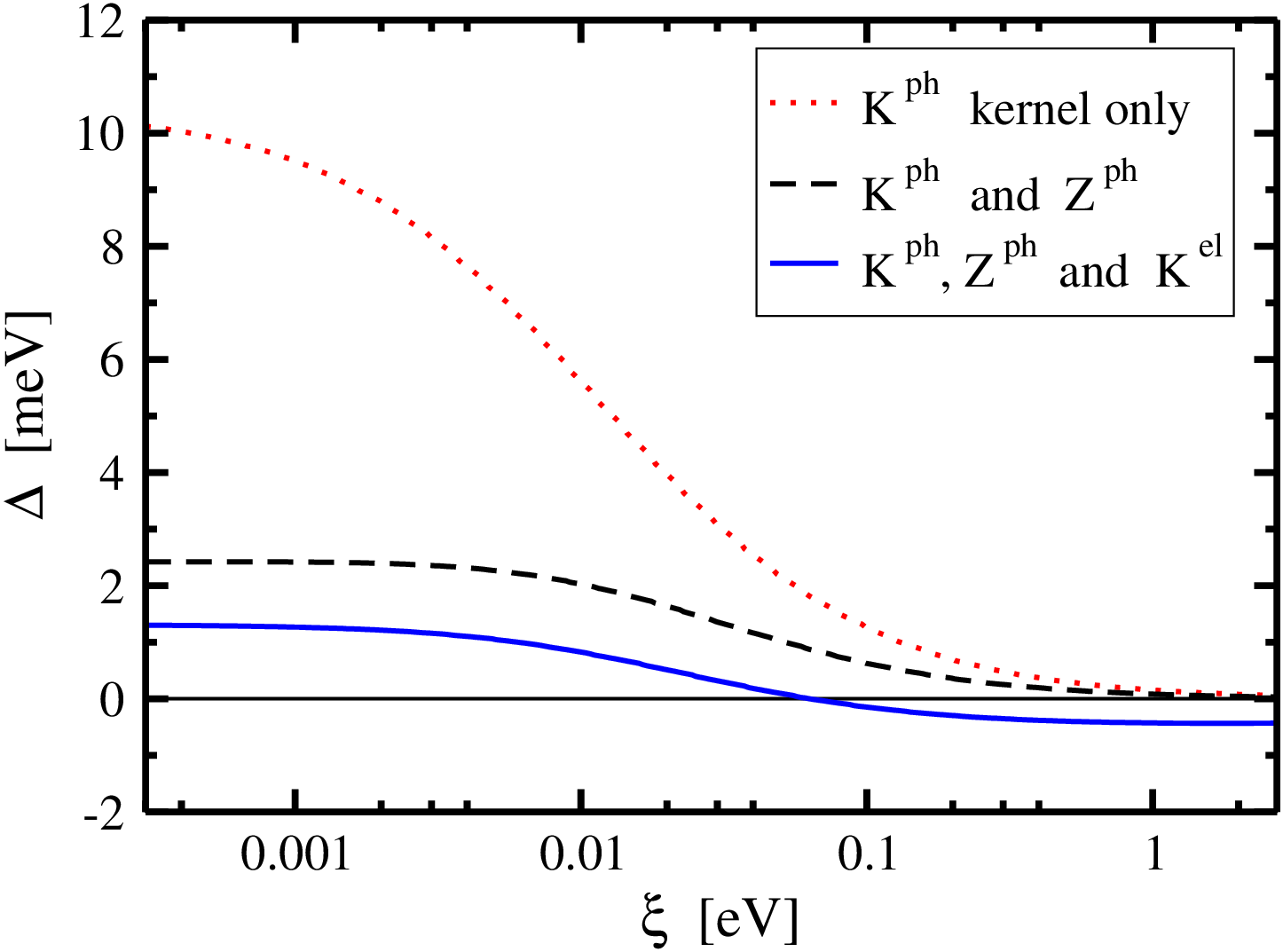}
\end{center}
\caption{(Color online)
Gap function versus energy at $T=0.01$\,K for Pb, including the different contributions to the
gap equation.}
\label{compacdcdc}
\end{figure}
It is instructive to visualize how the different terms entering the gap equation combine to
yield the values listed in Tab.~\ref{tbresfr}: The phononic term ${\cal K}^\text{ph}$ is negative,
and is the responsible for the superconducting state in the simple metals we studied; 
${\cal Z}^\text{ph}$ gives a repulsive contribution, whose relative importance increases
for the strong coupling materials; finally, the electronic Coulomb repulsion leads to
a large cancellation between constructive and destructive interference effects. To better
understand these effects we plot, in Fig.~\ref{compacdcdc}, the gap function at zero temperature
of Pb obtained through the solution of the gap equation~(\ref{eq:gap_E}) including only
${\cal K}^\text{ph}$, the two phononic contributions, or all three phononic and Coulomb 
contributions. Clearly, the value of the gap is the result of a subtle interplay of opposite 
contributions, each one considerably larger than the gap itself. 

\begin{figure*}
\begin{center}
  \includegraphics[width=0.85\columnwidth,clip]{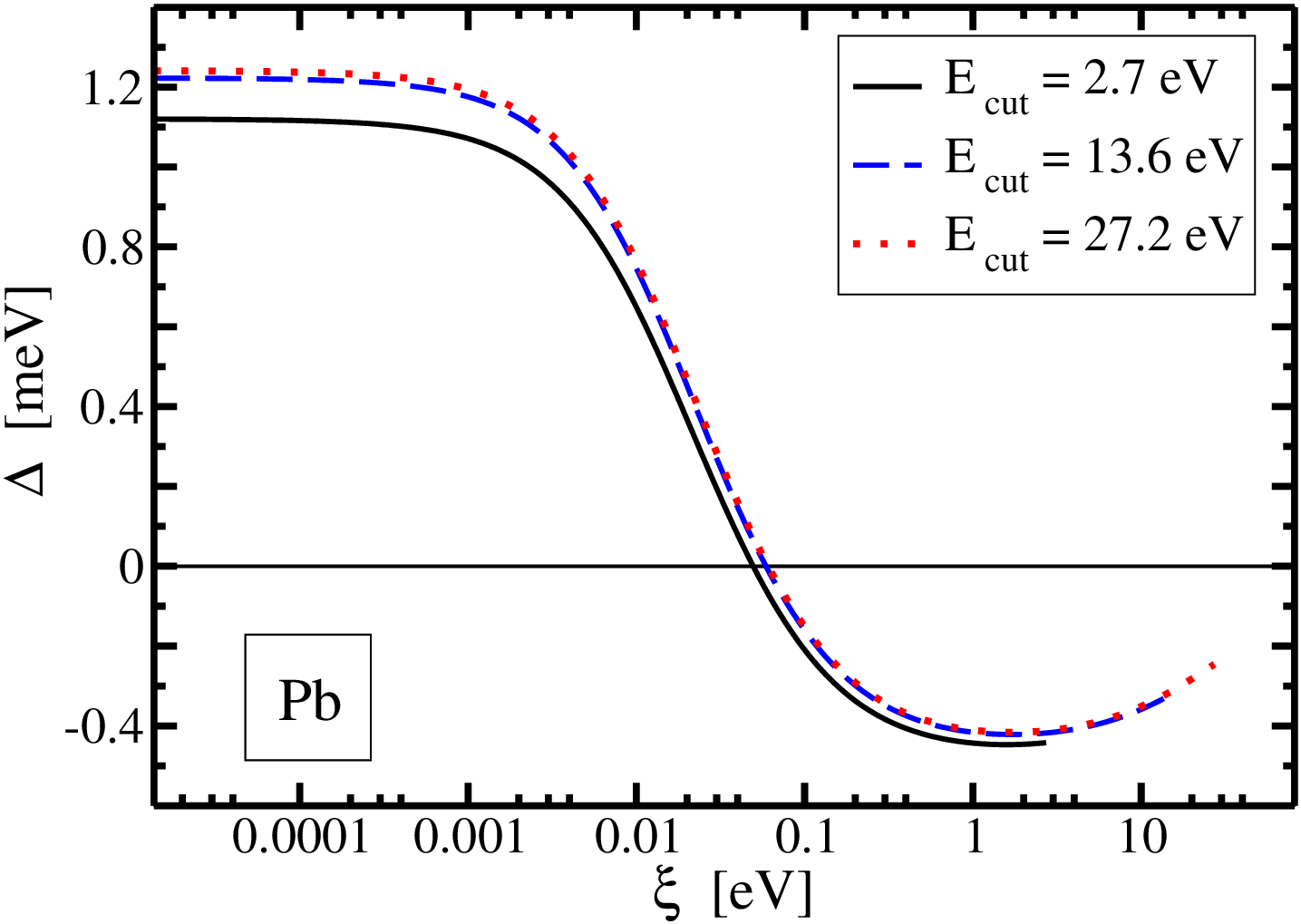}
  \qquad
  \includegraphics[width=0.85\columnwidth,clip]{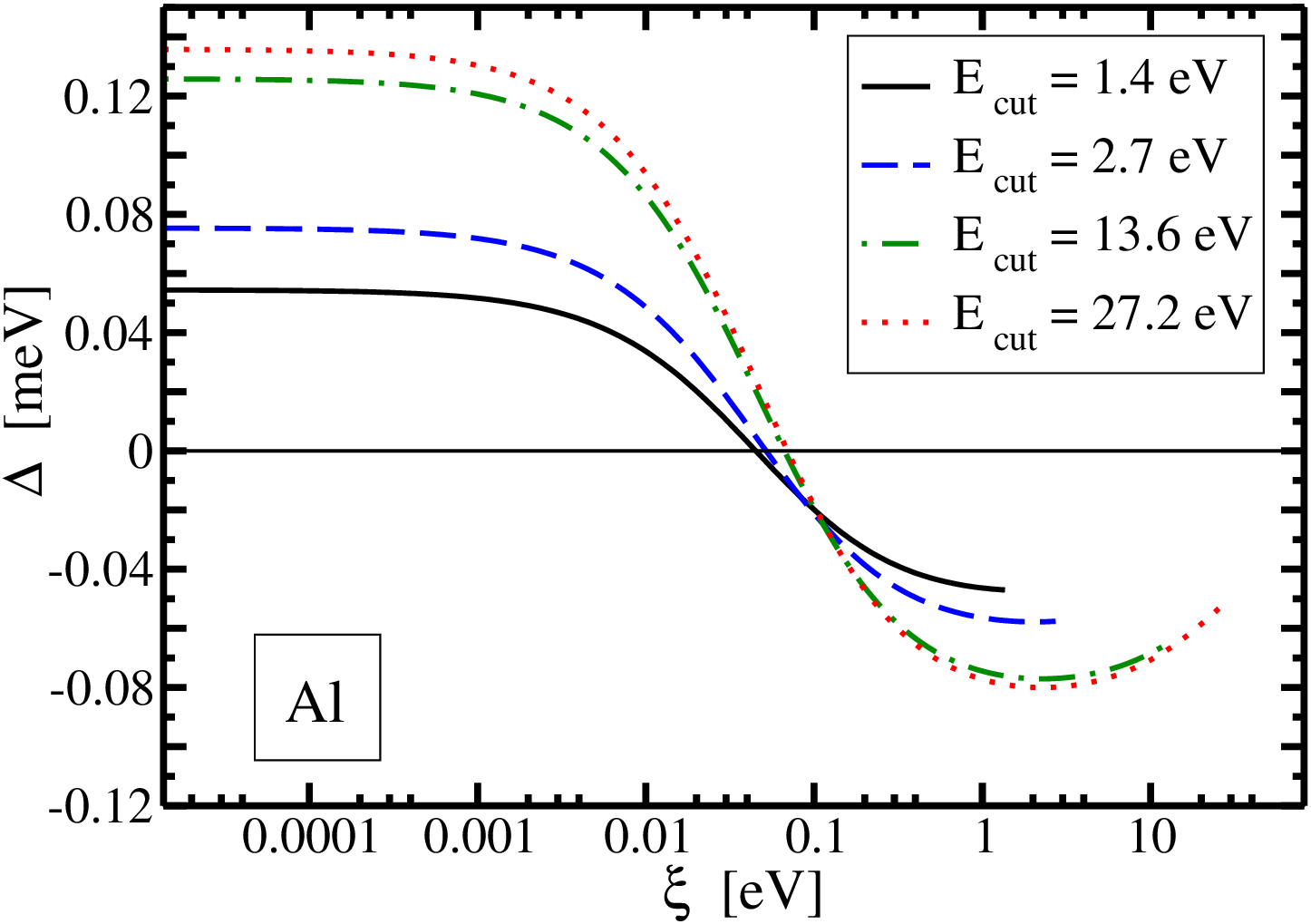}
\end{center}
\caption{(Color online)
Convergence of the calculated $T=0.01$~K gap as a function of the energy cutoff, for 
  Pb and Al.
  \label{conv_wind}}
\end{figure*}
Another interesting point regards the convergence of the gap function with the energy cutoff.
This convergence can be observed in Fig.~\ref{conv_wind}, where we plot the gap for Pb and Al
calculated within the \LDAzero\ approach. The different materials behave differently, with a 
faster convergence for the material with stronger electron-phonon coupling. However, even for Pb, 
an energy cutoff of at least 10\,eV was necessary to achieve convergence. It is a key feature
of our approach that the matrix elements of the screened Coulomb interaction are used in the kernel
of the gap equation up to very high energies. It is this feature that allows for the description
of non-trivial, material-specific effects. Traditionally, the use of this large energy window
is avoided by rescaling the Fermi-surface average, $\mu$, of these matrix elements. This 
rescaling leads to the the Morel-Anderson pseudopotential $\mu^*$. While  $\mu^*$ is an 
ingenious concept to capture the essential physics of the Coulomb repulsion in simple 
superconductors, it is not sufficient to treat more complex materials such as, e.g., $\rm MgB_2$.    
The detailed analysis carried out in Ref.~\onlinecite{MgB2-PRL} shows how  
the actual value of matrix elements of the screened Coulomb interaction, computed w.r.t.
Bloch states of different orbital character ($\sigma$ and $\pi$), leads to non-trivial effects, 
determined by the specific character of the orbitals.
This indicates that, contrary to common wisdom,
superconducting properties are not exclusively determined by a small region
around the Fermi level. 
Only the contributions from states higher than 20--30\,eV are essentially negligible, due to the asymptotic decay of the 
Coulomb term~(\ref{eq:vTF_e}).

\begin{figure*}
\begin{center}
  \includegraphics[width=0.85\columnwidth,clip]{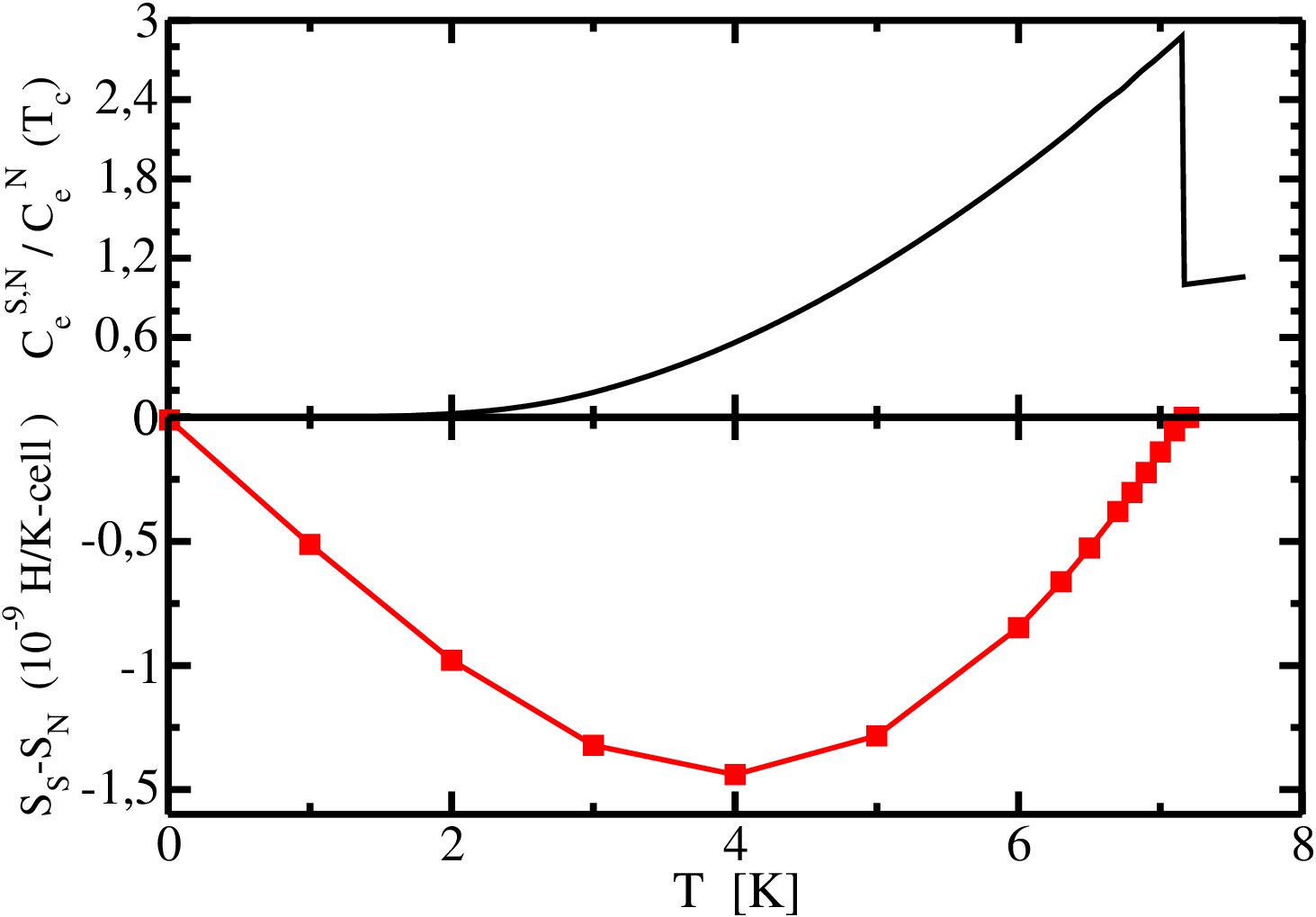}
  \qquad
  \includegraphics[width=0.85\columnwidth,clip]{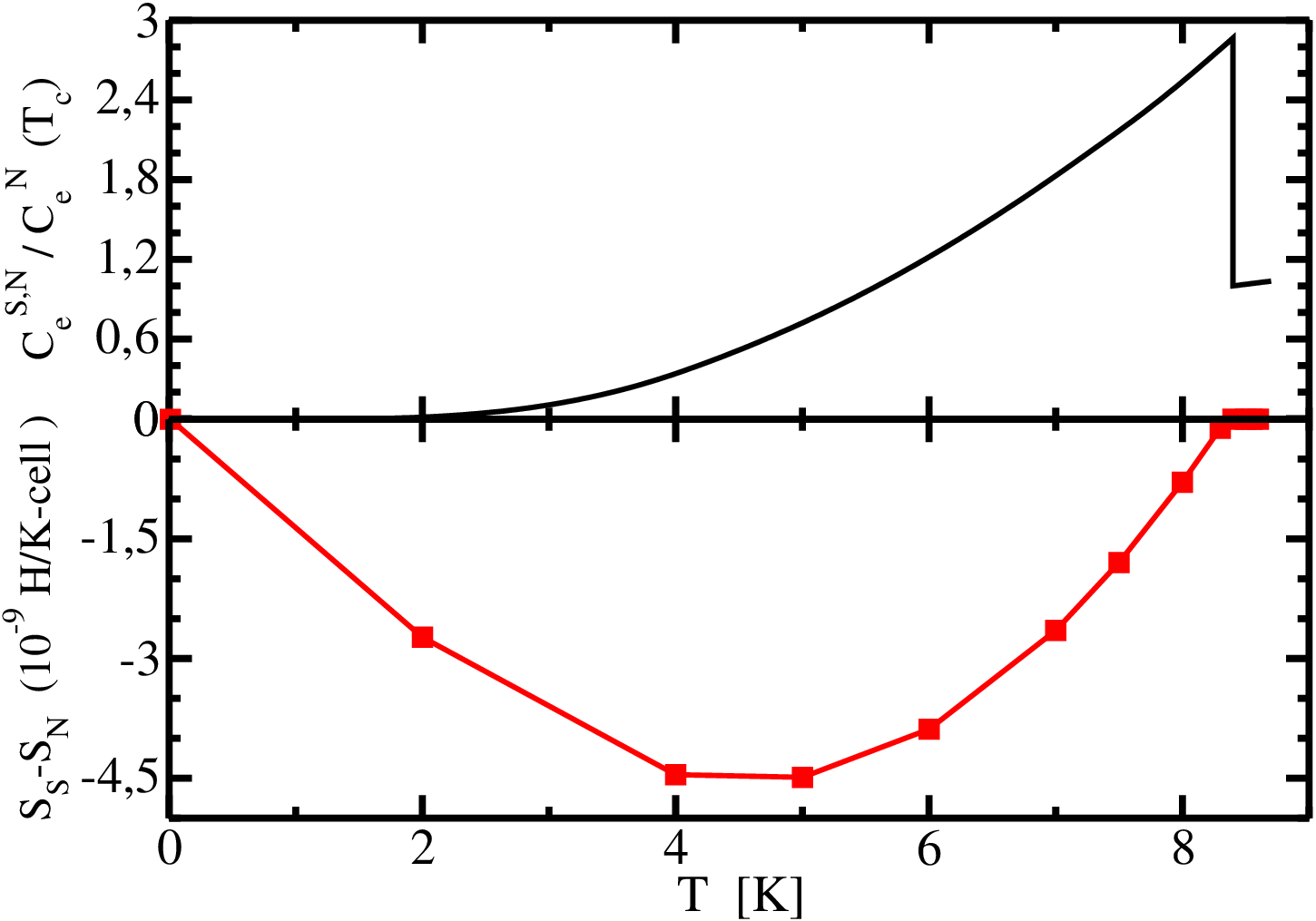}
\end{center}
\caption{(Color online) Thermodynamic functions for Pb (left panels) and Nb (right panels). 
  Upper panels: ratio between the electronic specific heat ratio
  in the normal (N) or superconducting (S) states and the specific heat at $T_\text{c}$ 
  ($C_\text{e}^\text{S,N}(T)/C_{e}^\text{N}(T_\text{c})$). Lower panels: difference of entropy 
  between the superconducting and the normal states $\Delta S=S_\text{S}-S_\text{N}$.
  \label{pb_nb_termo}}
\end{figure*}
After solving the gap equation, it is straightforward to calculate thermodynamic 
functions, such as the electronic entropy of the Kohn-Sham system
\begin{multline}
  \label{entros}
  S_\KS=-2 k_\text{B} \sum_\nk\Big\{f_\beta(E_\nk)\log\left[f_\beta(E_\nk)\right] \\
  +\left[1-f_\beta(E_\nk)\right]\log\left[1-f_\beta(E_\nk)\right]\Big\}
  \,,
\end{multline}
where $k_\text{B}$ is the Boltzmann constant. In Fig.~\ref{pb_nb_termo} we depict the
difference of entropy between the superconducting and the normal states $\Delta S=S_\text{S}-S_\text{N}$
for Pb and Nb. This quantity has the expected temperature dependence, going smoothly to zero at $T_\text{c}$. 
This indicates the stability of our calculations even close to the transition temperature where the gap becomes 
very small.
In the same figure we also depict the normalized specific heat $C_\text{e}^\text{S,N}(T)/C_{e}^\text{N}(T_\text{c})$. 
The specific heat is obtained by evaluating numerically the temperature derivative of the entropy. Despite the
numerical uncertainty associated to the calculation of the derivative, the curves of the specific heat are
quite stable. The discontinuities of the specific heat at $T_\text{c}$ obtained within the \DFT\ approach are
shown in Table~\ref{tbther}. Our results are in quite good agreement with experiment: While for
Al and Ta we confirm the BCS value found in experiments, we reproduce the strong coupling value of Nb.
For Pb the agreement with experiment is worse, but still within acceptable margins.

\begin{table}
\begin{center}
\begin{tabular}{ccc}
     &  Theory &   Experiment \\ \hline
Pb   &        2.93     &          3.57-3.71 \\
Nb   &        2.87     &          2.8-3.07  \\
Ta   &        2.64     &          2.63      \\
Al   &        2.46     &          2.43      \\
\end{tabular}
\caption{Normalized electronic specific heat $C_\text{e}^\text{S}(T_\text{c})/C_\text{e}^\text{N}(T_\text{c})$, as computed from
the \DFT\ approach.} 
\label{tbther} 
\end{center}
\end{table}

It is clear that any theory that aims at describing the superconducting state has to include retardation
effects. While the main equation of our density functional formalism, the gap equation~(\ref{eq:gap}), 
has the form of a static equation, retardation enters through the electron-phonon part of the
exchange-correlation potential. To further demonstrate that our theory takes retardation effects properly 
into account, we calculated the isotope effect coefficient $\alpha$. In fact, if retardation effects were not
included, $\alpha$ would be equal to the BCS value $\alpha=0.5$. For a mono-atomic solid, the Eliashberg
function $\alpha^2F(\Omega)$ for materials with different isotopic masses can be obtained by
rescaling the phonon frequencies with the square root of the nuclear mass $M$. Note that only the
frequencies are rescaled, while the total electron-phonon coupling constant $\lambda$ remains unchanged.
Within the \LDAzero\ approach, we computed the isotope effect of the gap function. We obtained the values 
0.37 and 0.47 for Mo and Pb respectively, which can be compared to the corresponding experimental values, 
$\alpha=0.33$ and $\alpha=0.47$. This agreement with experiment, in the presence a  significant deviation
from the BCS value, proves, in our opinion, that retardation effects are properly taken into account in
our calculations.

\begin{figure}
\begin{center}
\includegraphics[width=0.85\columnwidth,clip]{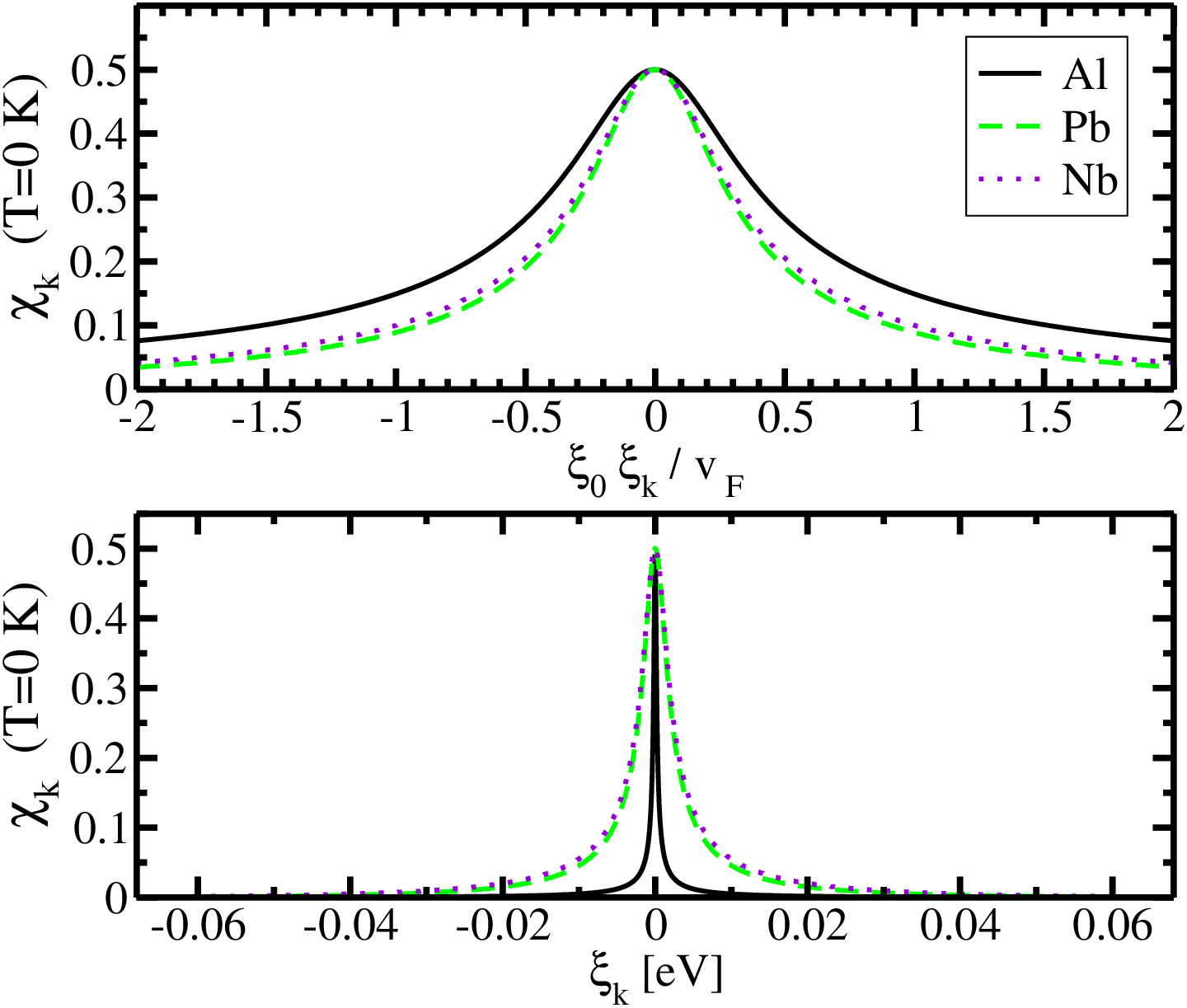}
\end{center}
\caption{(Color online) Order parameter for Al, Pb, and Nb represented on the basis of Bloch orbitals 
  as a function of momentum (upper panel) and energy (lower panel). The momentum was rescaled 
  with the experimental coherence lengths $\xi_0=$ 1600, 90, and 40\,nm for Al, Pb and Nb respectively.
  \label{ukvk}}
\end{figure}
To conclude the presentation of our results we show, in Fig.~\ref{ukvk} the order parameter represented
on the basis of Bloch orbitals. The lower panel of the figure depicts the order parameter for
Al, Pb and Nb as a function of the energy of the corresponding Bloch state. As expected,
$\chi_{\nk}$ is very localized in a small energy window around the Fermi level. The spread of
the order parameter in $\bk$-space is related to the inverse of the coherence length of the 
superconductor. This can be visualized by rescaling the relative energy $\xi$ by
$\xi_0/\hbar \bv_\text{F}$, where $\xi_0$ is the experimental coherence length of the material and
$\hbar \bv_\text{F}=\nabla_{\bk} \xik$ is the Fermi velocity. The resulting curves depict
the spread in $\bk$-space of the wave-packets describing the Cooper pairs in units of the
experimental coherence length. As expected, the curves for Al, Pb, and Nb are quite similar to each other, 
and have a width comparable to unity. We can therefore conclude that not only the maximum gap value but 
also its overall energy dependence are described accurately in our approach.

\section{Conclusions}
\label{sec:concl}

In this work we present the first full-scale application of the ab-initio theory for superconductivity,
which was developed in the preceeding paper (I).
Superconducting properties of simple conventional superconductors are computed without any experimental
input. In this way, we were able to test the theory developed in I and to assess the quality
of the functionals proposed. It turns out that the different proposed functionals lead to results
which are in good agreement with each other for the simple metals studied. The agreement is better
when the normal-state Bloch functions are not too strongly localized (as, e.g., the $sp$-orbitals in
metals in contrast to the more localized $d$ states of the transition elements). 
The most important result is that the calculated transition temperatures
and superconducting gaps are also in
good agreement with experimental values. The largest deviations from the experimental results
are found for the elements in the weak coupling limit with Mo being the most pronounced example.
We emphasize, however, that the agreement of our results with experiment, without making use of
any adjustable parameter, is unprecedented in the field.
Furthermore, we also obtained other quantities such as the entropy and the specific heat.
Our approach reproduces the correct values for the discontinuity of the specific heat at $T_\text{c}$
even in the strong coupling regime. Finally, we calculated the isotope effect for Mo and Pb,
achieving again rather good agreement with experiment. These results clearly show that retardation
effects are correctly described by the theory. Our calculations demonstrate that, as far as conventional
superconductivity is concerned, the ab-initio prediction of superconducting properties is feasible.

\begin{acknowledgments}
S.M. would like to thank A. Baldereschi, G. Cappellini, and G. Satta for useful discussions.
This work was partially supported by the INFM through the Advanced Research Project (PRA) UMBRA,   
by a supercomputing grant at Cineca 
(Bologna, Italy) through the Istituto Nazionale di Fisica della Materia (INFM), and by the 
Italian Ministery of Education, through a 2004 PRIN project. Financial support by the 
Deutsche Forschungsgemeischaft, by the EXC!TiNG 
Research and Training Network of the European Union, and by the NANOQUANTA Network of
Excellence is gratefully acknowledged. Part of the calculations were performed at the
Hochleistungsrechner Nord (HLRN).
\end{acknowledgments}



\end{document}